\documentclass[twocolumn]{aastex631}

\newcommand{\zem}{{\ifmmode{z_{\rm em}}\else{$z_{\rm em}$}\fi}}
\newcommand{\kms}{{\ifmmode{{\rm km~s}^{-1}}\else{km~s$^{-1}$}\fi}}
\newcommand{\cmm}{{\ifmmode{{\rm cm}^{-2}}\else{cm$^{-2}$}\fi}}
\newcommand{\lya}{Ly$\alpha$}
\newcommand{\lyb}{Ly$\beta$}

\newcounter{species} 
\def\ion#1#2{\setcounter{species}{#2}#1$\;${\scriptsize\Roman{species}}\relax}

\shorttitle{TPE around BAL quasars}
\shortauthors{Misawa et al.}

\graphicspath{{./}{figure/}}

\begin{document}

\title{Exploratory Study of Transverse Proximity Effect around BAL
  Quasars}

\correspondingauthor{Toru Misawa}
\email{misawatr@shinshu-u.ac.jp}

\author[0000-0002-5464-9943]{Toru Misawa}
\affil{School of General Education, Shinshu University, 3-1-1 Asahi,
  Matsumoto, Nagano 390-8621, Japan}

\author[0000-0002-2134-2902]{Rikako Ishimoto}
\affil{Department of Astronomy, Graduate School of Science, The
  University of Tokyo, 7-3-1 Hongo, Bunkyo, Tokyo 113-0033, Japan}

\author{Satoshi Kobu}
\affil{Department of Physics, Faculty of Science, Shinshu University,
  3-1-1 Asahi, Matsumoto, Nagano 390-8621, Japan}

\author[0000-0003-3954-4219]{Nobunari Kashikawa}
\affil{Department of Astronomy, Graduate School of Science, The
  University of Tokyo, 7-3-1 Hongo, Bunkyo, Tokyo 113-0033, Japan}
\affil{Research Center for the Early Universe, The University of
  Tokyo, 7-3-1 Hongo, Bunkyo-ku, Tokyo 113-0033, Japan}

\author[0000-0003-3466-3876]{Katsuya Okoshi}
\affil{Institute of Arts and Sciences, Tokyo University of Science,
    6-3-1, Niijyuku, Katsushika, Tokyo 125-8585, Japan}

\author[0000-0002-5197-8944]{Akatoki Noboriguchi}
\affil{School of General Education, Shinshu University, 3-1-1 Asahi,
  Matsumoto, Nagano 390-8621, Japan}

\author[0000-0001-7825-0075]{Malte Schramm}
\affil{Graduate school of Science and Engineering, Saitama Univ. 255
  Shimo-Okubo, Sakura-ku, Saitama City, Saitama 338-8570, Japan}

\author{Qiang Liu}
\affil{Department of Physics, Faculty of Science, Shinshu University,
  3-1-1 Asahi, Matsumoto, Nagano 390-8621, Japan}

\begin{abstract}
We aim to find out the reason why there exists an anisotropic
\ion{H}{1} absorption around quasars; i.e., the environments around
quasars are highly biased toward producing strong \ion{H}{1}
absorption in transverse direction while there exists a significant
deficit of \ion{H}{1} absorption within a few Mpc of quasars along
line-of-sight. The most plausible explanation for this opposite trend
is that the transverse direction is shadowed from the quasar UV
radiation due to dust torus. However, a critical weakness of this idea
is that we have no information on inclination angle of our sightline
relative to the torus. In this study, we examine environments of
quasars with broad absorption troughs in their spectra (i.e., BAL
quasars) because it is widely believed that BAL troughs are observed
if the central continuum is viewed from the side through their
powerful outflows near the dust torus.  With closely separated 12
projected quasar pairs at different redshift with separation angle of
$\theta$$<$120$^{\prime\prime}$, we examine \ion{H}{1} absorption at
foreground BAL quasars in spectra of background quasars.  We confirm
there exist optically thick gas around two of 12 BAL quasars, and that
the mean \ion{H}{1} absorption strength is EW$_{\rm
  rest}$$\sim$1\AA. These are consistent to the past results around
non-BAL quasars, although not statistically significant. However, the
origins of optically thick \ion{H}{1} absorbers around BAL and non-BAL
quasars could be different since their column densities are different
by $\sim$3 orders of magnitude. The larger sample would be required
for narrowing down possible scenarios for the anisotropic \ion{H}{1}
absorption around quasars.
\end{abstract}


\section{Introduction} \label{sec:intro}
Quasars have potential to influence the evolution of their host
galaxies as well as their environments by ejection of mass, energy,
and angular momentum through outflowing winds and energetic radiation
from accretion disks (i.e., feedback effect). The radiation effect has
been partially confirmed as the relative lack of \ion{H}{1} absorbers
in circum-galactic medium (CGM) and inter-galactic medium (IGM) in the
vicinity of quasars along the line of sight within a few Mpc of the
quasar redshift, produced by the enhancement of local radiation field
by quasars (known as Line-of-sight Proximity effect; LPE)
\citep{baj88,sco00}.  Indeed, the enhancement in the radiation field
by the quasar relative to the extragalactic UV background is, for
example, a factor of $\sim$1000 at a proper scale of 100~kpc ($\sim$10
at 1~Mpc) from quasars with a typical bolometric luminosity of $\log
L_{\rm bol}$/(ergs~s$^{-1}$) $\sim$ 46 at $z$ $\sim$ 2.5.  If the
quasars emit UV radiation isotropically, we should detect the same
trend (i.e., a deficit in \ion{H}{1} absorption) in {\it transverse}
direction, too (so-called transverse proximity effect; TPE). However,
the TPE has not been detected in spite of many attempts (e.g.,
\citealt{cro04} and references therein).

\citet{hen06} carried out an intensive quasar survey to examine
environments around quasars in transverse direction using quasar
pairs.  As a part of this ``quasars probing quasars (QPQ)'' project
whose targets are mainly Type~1 quasars, \citet{pro13} collected 650
projected quasar pairs from SDSS/BOSS quasar catalog with physical
separation of $\lesssim$ 1~Mpc at $z$ $\sim$ 2.5.  They revealed that
i) absorption strength (i.e., rest-frame equivalent width; EW$_{\rm
  rest}$) of \ion{H}{1} increases monotonically with decreasing a
transverse distance ($R_{\perp}$) from quasars by EW$_{\rm rest}$ =
2.3$\AA$($R_{\perp}$/100~kpc)$^{-0.46}$, and ii) a detection rate of
optically thick \ion{H}{1} absorbers with $\log N_{\rm
  HI}$/(cm$^{-2}$) $\geq$ 17.2 is very high ($\sim$60\%\ at
$R_{\perp}$ $<$ 100~kpc and $\sim$20\%\ at $R_{\perp}$ $\simeq$ 1~Mpc,
respectively), while the detection rate near the quasars along the
line of sight is only $\sim$5\%, close to the random distribution.
The discrepancy still remains even after correcting over density
effect \citep{jal19}.  Thus, regions transverse to quasars exhibit
{\it enhanced} \ion{H}{1} absorption in contrast to the measurements
along the line of sight.

This opposite trend implies that the physical condition of \ion{H}{1}
absorbers is highly anisotropic around Type~1 quasars.  The most
plausible explanation is that the transverse direction is shadowed
from the quasar UV radiation due to obscuration effect (an anisotropic
scenario; \citealt{pro13,jal19}).  An anisotropic radiation can be
realized by AGN unification models where the black hole is obscured by
a dust torus (e.g., \citealt{ant93,elv00}).  Indeed, studies of Type~2
quasars and the X-ray background suggests that quasars have $\sim$30\%
of the solid angle obscured \citep{ued03,tre05}.  This scenario is
also supported by the negative results for detecting Ly$\alpha$
fluorescent emission from optically thick \ion{H}{1} absorbers near
the quasars in transverse direction whose fluorescent surface
brightness exceeds the detection limit if we assume isotropic
radiation \citep{hen13}.  Thus, the anisotropic scenario is promising,
but with a single critical weakness remains; we have no information on
viewing angle (i.e., inclination angle) of our sightline relative to
the dust torus.

Broad absorption troughs with FWHM $\geq$ 2000~km/s (i.e., broad
absorption lines; BALs) are present in spectra of about 15 -- 40\%\ of
optically-selected quasars (e.g., \citealt{wey91,kni08,all11}).  There
are two possible scenarios to explain the existence of BAL
features. The first one is an orientation scenario; BALs are observed
if normal quasars are viewed along a particular line of sight, almost
parallel to the radiatively driven equatorial outflow wind whose
inclination angle relative to the rotation axis of an accretion disk
is large and close to the direction of (but not fully obscured by)
dust torus (e.g., \citealt{mur95,pro00}).  The another possibility is
an evolution scenario; BAL quasars (especially low-ionization BALs
(LoBALs) with absorption lines of low-ionization species like
\ion{Mg}{2}) could represent an early stage in the lifetime of the
quasar. At this stage, an extreme starburst approaches its ends and
the surrounding dust cocoon are being blown away. Indeed, a
significant fraction of them are dust-reddened quasars
\citep{far07,urr09} and star formation rate depends on the existence
of (Lo)BALs \citep{che22}.

\citet{pro13} and \citet{jal19} excluded BAL quasars from their
samples.  Thus, the environment around BAL quasars (i.e., our
sightline is close to edge-on) is expected to be different from those
around non-BAL quasars (close to face-on). Here, we emphasize that we
cannot examine the LPE around BAL quasars because broad and strong BAL
features prevent us from measuring \ion{H}{1} absorption amplitude
adequately near the quasars.

If we find that an \ion{H}{1} absorption around BAL quasars in the
transverse direction is weaker than those expected around non-BAL
quasars, it would lead to some important results: i) quasar radiation
is anisotropic as expected due to the dust torus, ii) BAL winds are
driven into a generally equatorial direction (i.e., closer to dust
torus) as expected, and iii) the detectability of BAL features mainly
depends on our viewing angles to the BAL wind (i.e., the orientation
scenario) with the evolution scenario as a secondary effect.  On the
other hand, if an \ion{H}{1} absorption around BAL quasars in the
transverse direction is at the same level as seen around non-BAL
quasars, we can confirm that i) we always observe Type~1 quasars
regardless of the existence of BAL profiles, and/or that ii) we cannot
use BAL quasars as an indicator that they are being observed from a
direction close to the side (i.e., the evolution scenario is more
applicable to BAL features; e.g., \citealt{lip06}).

In this study, we perform the survey for BAL quasars that were
excluded in the past studies.  We examine the proximity effect around
quasars in transverse direction using BAL quasars and compare them to
non-BAL quasars from the literature, and locate the origin of the
anisotropically distributed \ion{H}{1} absorbers.  We describe target
selection and the observations in \S2, and the data analysis in
\S3. The results and discussion are presented in \S4 and \S5.  We
close by summarizing our results in \S6.  Throughout the paper, we use
a cosmology with $H_{0}$=70~\kms~Mpc$^{-1}$, $\Omega_{m}$=0.26, and
$\Omega_{\Lambda}$=0.74 matching to those used in \citet{pro13}.

\section{Sample and Observations} \label{sec:obs}

\subsection{Sample Selection} \label{sec:sample}
We searched the quasar catalogs of SDSS/BOSS DR16 \citep{lyk20} for
quasar pairs that satisfy the criteria below: 1) an emission redshift
of a foreground (f/g) quasar ($z_{\rm f/g}$) is greater than 2.0 so
that we can detect \ion{H}{1} absorption lines at observed wavelength
of $\lambda_{\rm obs}$ $\gtrsim$ 3650\AA\ in SDSS/BOSS spectra of a
background (b/g) quasar, 2) a separation angle between quasars is
$\theta < 120^{\prime\prime}$ which corresponds to $R_{\perp}$ $\lesssim$
1~Mpc at $z$ $\sim$ 2.5 within which the radiation from the quasar
dominates the UV background (i.e., the ionization condition of the CGM
and IGM strongly depends on the existence of the dust torus of a f/g
quasar), 3) a BAL profile is present in the spectrum of a f/g quasar
(i.e., Balnicity Index\footnote{Balnicity Index was originally
  introduced by \citet{wey91} as
\begin{equation}
BI = \int_{3000}^{25000}\left(1 - \frac{f(-v)}{0.9}\right)Cdv,
\end{equation}
where $f(-v)$ is the continuum-normalized flux at a velocity of $v$
(in \kms) from \ion{C}{4} or \ion{Si}{4} emission-line center, $C$ is
a constant equal to 1 in regions where $f(-v)$ has been continuously
less than 0.9 for at least 2000~\kms\ (otherwise, $C$ = 0).} (BI) $>$
0) while a b/g quasar has no remarkable absorption feature (i.e., BI =
0 and Absorption Index\footnote{Absorption Index is computed by the
  same equation as BI, but the integration range is slightly different
  as defined by \citet{hal02}.} (AI) = 0)\footnote{We used the values
  of BI and AI in \citet{lyk20}, but we also visually checked the
  existence of strong absorption lines ourselves.}, 4) a radial
velocity difference between f/g and b/g quasars, $\Delta v$~($z_{\rm
  b/g} - z_{\rm f/g}$), is larger than 5,000~km/s to avoid the
line-of-sight proximity effect of a b/g quasar, 5) \ion{H}{1}
Ly$\alpha$ absorption lines at $z_{\rm f/g}$ do not blend with
Ly$\beta$ or higher order Lyman series lines at $z$ $\gtrsim$ $z_{\rm
  f/g}$ (i.e., (1+$z_{\rm f/g}$)$\times$1216 should be larger than
(1+$z_{\rm b/g}$)$\times$1026)), and 6) $g$-band magnitude of both f/g
and b/g quasars are bright enough ($g$ $\lesssim$ 20.0) to acquire a
high-quality spectrum.

\begin{figure}[ht!]
  \begin{center}
    \includegraphics[width=8cm,angle=0]{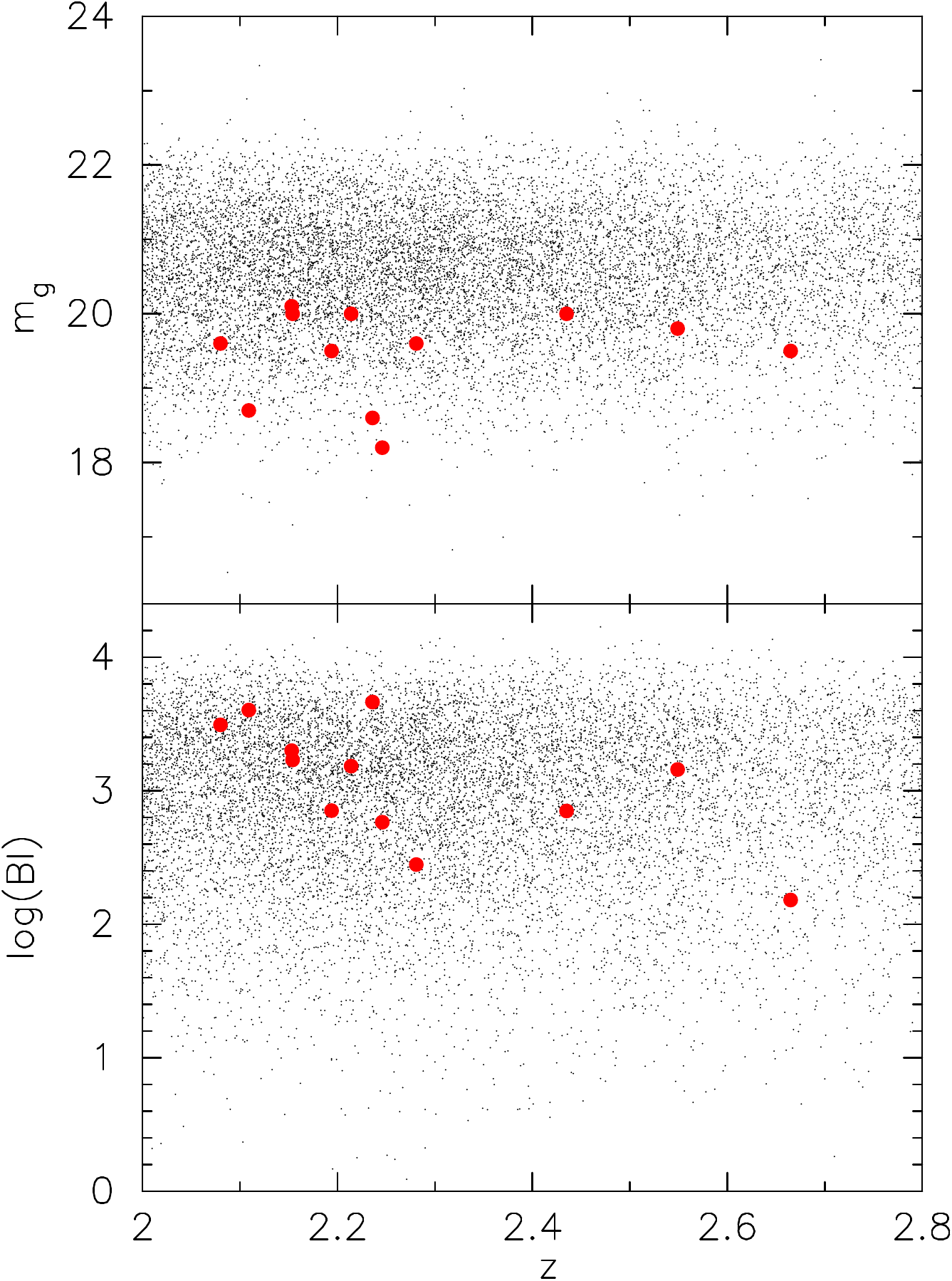}
  \end{center}  
  \caption{$g$-band magnitude ($m_{\rm g}$) and Balnicity index (BI)
    as a function of redshift of our 12 foreground quasars (red dotted
    circles) and all BAL quasars (black dots) from \citet{lyk20}. In
    the redshift range shown, a flux at 1325\AA\ in the rest-frame
    (which is used to estimate the specific flux in
    Section~\ref{sec:enhance}) is redshifted into the wavelength range
    of g-band filter.\label{fig:sample}}
\end{figure}

Among 750,414 quasars in \citet{lyk20}, we found 12 projected quasar
pairs\footnote{Before placing the final criterion on $g$-band
  magnitude, we have 337 projected quasar pairs that would be possible
  targets for future observations.} satisfying all the criteria above
as listed in Table~\ref{tab:qsos}. Our sample quasars are well
representative with respect to the global BAL quasar population, as
compared in Figure~\ref{fig:sample}, except that they are biased
toward brighter ones (due to our selection criterion).  We will refer
to them as PQ1 through PQ12 according to their coordinates.  Some of
them have been observed more than once in the Sloan Digital Sky Survey
(SDSS)-I/II/III/IV as shown in Table~\ref{tab:obs}.

%%% Table 1 %%%
\begin{deluxetable*}{ccccccccc}
\tablecaption{Sample Quasars\label{tab:qsos}}
\tablewidth{0pt}
\tablehead{
\colhead{ID}             &
\colhead{f/g or b/g}     &
\colhead{quasar}         &
\colhead{\zem$^a$}       &
\colhead{$m_{\rm g}$}      &
\colhead{$m_{\rm i}$}      &
\colhead{BI$^b$}         &
\colhead{AI$^c$}         &
\colhead{$\theta$$^d$}   \\
\colhead{}               &
\colhead{}               &
\colhead{}               &
\colhead{}               &
\colhead{(mag)}          &
\colhead{(mag)}          &
\colhead{(\kms)}         &
\colhead{(\kms)}         &
\colhead{(arcsec)}       
}
\startdata
PQ1  & f/g & SDSS~J003135.57+003421.2 & 2.236$\pm$0.003$^e$ & 18.6 & 18.4 & 4618 & 5602 &  62 \\
     & b/g & SDSS~J003138.18+003333.2 & 2.741$\pm$0.004 & 19.2 & 19.0 &    0 &    0 &     \\
PQ2  & f/g & SDSS~J021631.70+034009.9 & 2.109$\pm$0.003 & 18.7 & 18.1 & 4017 & 5440 &  25 \\
     & b/g & SDSS~J021630.65+034029.4 & 2.445$\pm$0.003 & 19.7 & 19.7 &    0 &    0 &     \\
PQ3  & f/g & SDSS~J092223.04+160531.4 & 2.281$\pm$0.003 & 19.6 & 19.4 &  280 & 1234 & 101 \\
     & b/g & SDSS~J092216.04+160526.4 & 2.373$\pm$0.003 & 18.1 & 18.0 &    0 &    0 &     \\
PQ4  & f/g & SDSS~J095612.57+370052.9 & 2.153$\pm$0.003 & 20.1 & 19.8 & 1997 & 2458 &  71 \\
     & b/g & SDSS~J095618.43+370106.2 & 2.686$\pm$0.004 & 20.0 & 19.8 &    0 &    0 &     \\
PQ5  & f/g & SDSS~J110532.92+515559.7 & 2.214$\pm$0.003 & 20.0 & 19.5 & 1527 & 2893 &  46 \\
     & b/g & SDSS~J110529.06+515530.3 & 2.439$\pm$0.003 & 19.9 & 19.7 &    0 &    0 &     \\
PQ6  & f/g & SDSS~J123042.00+363527.6 & 2.194$\pm$0.003 & 19.5 & 19.4 &  708 & 1004 &  91 \\
     & b/g & SDSS~J123049.33+363550.4 & 2.649$\pm$0.004 & 19.7 & 19.7 &    0 &    0 &     \\
PQ7  & f/g & SDSS~J124547.67+341822.6 & 2.080$\pm$0.003 & 19.6 & 19.2 & 3110 & 5358 &  73 \\
     & b/g & SDSS~J124546.58+341934.8 & 2.593$\pm$0.004 & 19.9 & 19.7 &    0 &    0 &     \\
PQ8  & f/g & SDSS~J135417.90+334859.8 & 2.246$\pm$0.003 & 18.2 & 18.0 &  580 &  985 & 109 \\
     & b/g & SDSS~J135414.28+334720.4 & 2.321$\pm$0.003 & 19.8 & 19.6 &    0 &    0 &     \\
PQ9  & f/g & SDSS~J144539.14+373336.0 & 2.665$\pm$0.004 & 19.5 & 19.4 &  152 &  489 &  69 \\
     & b/g & SDSS~J144544.90+373328.9 & 2.963$\pm$0.004 & 19.2 & 19.1 &    0 &    0 &     \\
PQ10 & f/g & SDSS~J150723.66+255216.6 & 2.549$\pm$0.004 & 19.8 & 19.6 & 1441 & 2504 &  92 \\
     & b/g & SDSS~J150717.25+255144.6 & 2.718$\pm$0.004 & 19.5 & 19.3 &    0 &    0 &     \\
PQ11 & f/g & SDSS~J224015.85+221230.2 & 2.435$\pm$0.003 & 20.0 & 19.8 &  704 & 2175 & 110 \\
     & b/g & SDSS~J224013.34+221414.1 & 2.675$\pm$0.004 & 19.0 & 18.9 &    0 &    0 &     \\
PQ12 & f/g & SDSS~J233011.09+254731.1 & 2.154$\pm$0.003 & 20.0 & 19.3 & 1699 & 2460 &  93 \\
     & b/g & SDSS~J233017.52+254803.0 & 2.297$\pm$0.003 & 19.2 & 19.1 &    0 &    0 &     \\
\enddata
\tablenotetext{a}{\ion{Mg}{2}-based emission redshift and its
  uncertainty (see Section~\ref{sec:zsys}).}
\tablenotetext{b}{Balnicity Index from \citet{lyk20}.}
\tablenotetext{c}{Absorption Index from \citet{lyk20}.}
\tablenotetext{d}{Angular separation between projected pair quasars in
  arcsec.}
\tablenotetext{e}{Emission redshift from \citet{fil13}.}
\end{deluxetable*}

%%% Table 2 %%%
\begin{deluxetable}{ccccccc}
\tablecaption{Observations of b/g Quasars \label{tab:obs}}
\tablewidth{0pt}
\tablehead{
\colhead{Target}                &
\colhead{Obs Date}              &
\colhead{Survey/Instrument$^a$} &
\colhead{S/N$^b$}               \\
\colhead{}                      &
\colhead{(mm-dd-yyyy)}          &
\colhead{}                      &
\colhead{(pix$^{-1}$)}       
}
\startdata
PQ1  b/g & 09-08-2010 & SDSS (4220-55447-0966) & 11 \\
         & 12-19-2001 & SDSS (0689-52262-0504) &  7 \\
         & 10-25-2017 & SDSS (9409-58051-0509) &  7 \\
PQ2  b/g & 11-06-2010 & SDSS (4264-55506-0464) &  6 \\	
PQ3  b/g & 03-12-2012 & SDSS (5307-55998-0198) & 25 \\
         & 03-24-2006 & SDSS (2440-53818-0452) & 17 \\
         & 02-27-2012 & SDSS (5305-55984-0944) & 18 \\
         & 01-30-2017 & SDSS (9581-57783-0704) & 21 \\
PQ4  b/g & 01-26-2011 & SDSS (4573-55587-0944) &  7 \\
PQ5  b/g & 04-03-2013 & SDSS (6706-56385-0074) &  8 \\
PQ6  b/g & 04-16-2010 & SDSS (3965-55302-0060) &  8 \\	
PQ7  b/g & 05-06-2010 & SDSS (3971-55322-0189) &  6 \\
PQ8  b/g & 03-19-2010 & SDSS (3861-55274-0718) &  9 \\
PQ9  b/g & 04-29-2012 & SDSS (5173-56046-0008) & 15 \\
         & 04-20-2004 & SDSS (1382-53115-0607) &  8 \\
         & 06-18-2021 & Subaru/FOCAS           & 22 \\
PQ10 b/g & 05-27-2012 & SDSS (6019-56074-0800) & 11 \\
         & 06-18-2021 & Subaru/FOCAS           & 42 \\
PQ11 b/g & 10-05-2012 & SDSS (6307-56205-0418) & 12 \\
         & 08-30-2005 & SDSS (2261-53612-0283) & 11 \\
         & 09-11-2012 & SDSS (6119-56181-0831) & 11 \\
         & 06-18-2021 & Subaru/FOCAS           & 28 \\
PQ12 b/g & 10-05-2013 & SDSS (6304-56570-0848) & 10 
\enddata
\tablenotetext{a}{Data source (SDSS or Subaru/FOCAS). In the case of
  SDSS, {\tt plate-MJD-fiber} information is added in parenthesis.}
\tablenotetext{b}{Signal to noise ratio at $\lambda_{\rm obs}$ $\sim$
  1215.67$\times$(1+$z_{\rm{f/g}}$) per pixel ($\Delta \lambda$ $\sim$
  1.0\AA\ and 0.75\AA\ for SDSS and Subaru/FOCAS spectra,
  respectively). The S/N ratio is estimated not in the observed (i.e.,
  absorbed) flux, but in the estimated continuum of the b/g quasars.}
\end{deluxetable}

%%% Figure 1 %%%

\subsection{Subaru/FOCAS Observations}
We additionally obtained spectra of b/g quasars of 3 relatively bright
quasar pairs (PQ9, PQ10, and PQ11) with Subaru/FOCAS on 2021 June 18,
using a similar resolution power ($R$ $\sim$ 2,000) to those of the
SDSS/BOSS spectra. These quasars have the highest redshift of f/g
quasars among our sample, and two of them (PQ10 and PQ11) have strong
\ion{H}{1} absorption systems (i.e., Damped \lya\ (DLA) systems) as
discovered by \citet{lyk20}.\footnote{They are identified at $z_{\rm
    DLA}$ = 2.539 and 2.443 in b/g quasars of PQ10 and PQ11,
  respectively.} Precise measurement of their \ion{H}{1} column
density is crucial for our analysis because one of our purposes is to
estimate the covering fraction of optically thick absorbers like DLAs
around the BAL quasars.  We observed them using a VPH450 grism with a
0.$^{\!\!\prime\prime}$6 slit width under clear and stable sky
conditions with the typical seeing size of
0.$^{\!\!\prime\prime}$7. The CCD was binned every 2 pixels in the
both spatial and dispersion directions (i.e., $\sim$1.28\AA\ per
pixel). We adopt 5 pixels as an extraction aperture.  Total exposure
time is 5400s, 5400s, and 8400s for b/g quasars of PQ9, PQ10, and
PQ11, respectively.

We reduced the data in a standard manner with the IRAF
software\footnote{IRAF is distributed by the National Optical
  Astronomy Observatories, which are operated by the Association of
  Universities for Research in Astronomy, Inc., under cooperative
  agreement with the National Science Foundation.}. Wavelength
calibration was performed using the spectrum of a Th-Ar lamp.  We
carried out the flux calibration of the spectra using the spectrum of
the spectrophotometric standard star Hz~44. The final S/N ratios of
the three quasars at $\lambda_{\rm obs}$ $\sim$
1215.67$\times$(1+$z_{\rm{f/g}}$) are improved from those of SDSS
spectra as shown in Table~\ref{tab:obs}.

\section{Analysis}

\subsection{Normalization of background quasar spectra}
\begin{figure}[ht!]
  \begin{center}
    \includegraphics[width=8cm,angle=0]{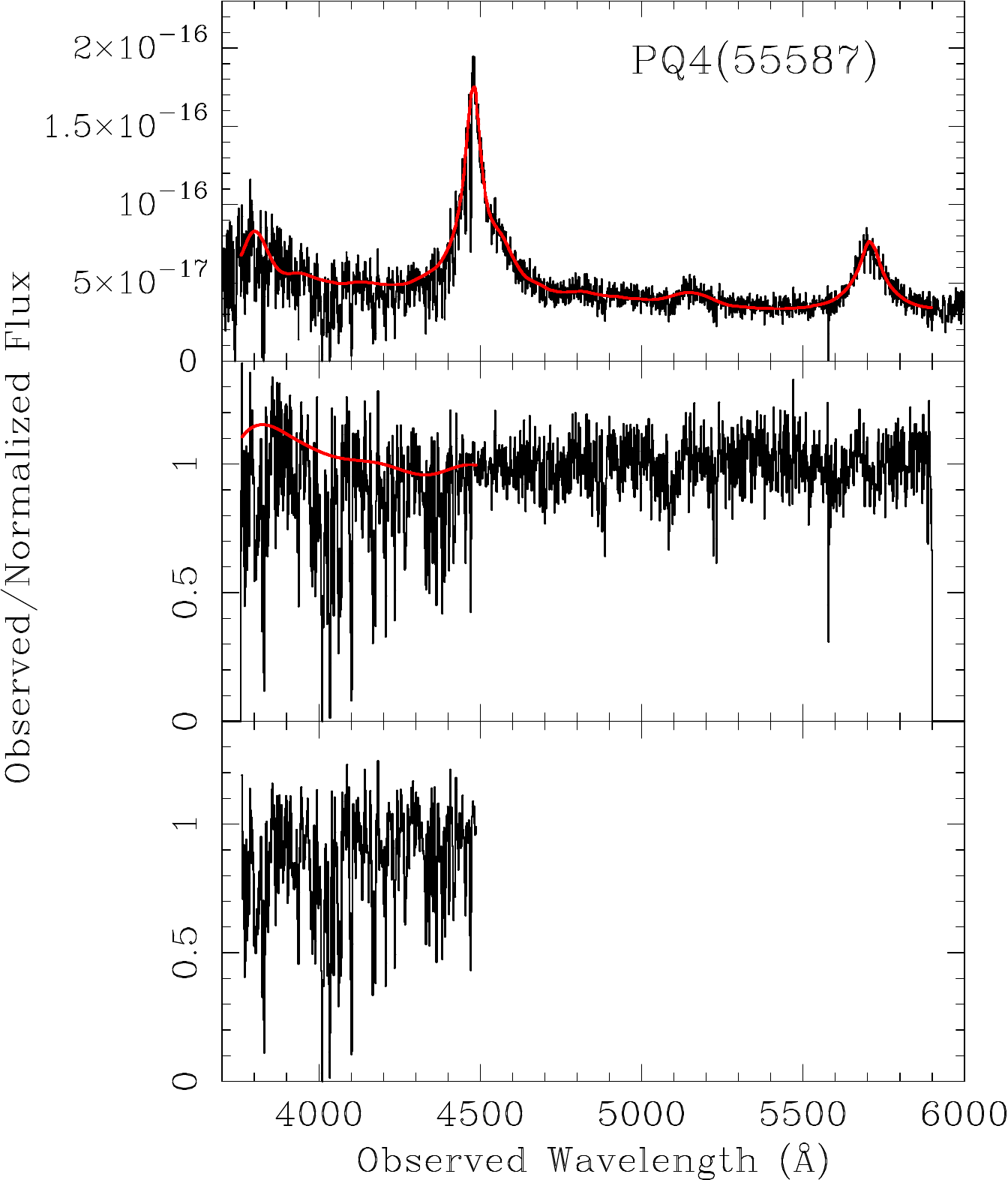}
  \end{center}  
  \caption{Normalization procedure for the SDSS spectrum (MJD:55587)
    of a b/g quasar of PQ4 as an example. Top panel: PCA-fit (red
    curve) to the observed spectrum (black histogram). Middle panel:
    fit model by a low-order spline function (red curve) to the
    PCA-normalized spectrum (black histogram). Bottom panel: final
    normalized spectrum whose transmitted mean flux is consistent with
    the expected value by \citet{fau08}.\label{fig:norm}}
\end{figure}

Before measuring \ion{H}{1} optical depth, we need to normalize
spectra of b/g quasars. However, wavelength regions shorter than the
\lya\ emission lines are severely affected by \lya\ forest, which
prevents us from directly fitting the continuum level. Therefore, we
estimate the continuum level of the quasar intrinsic spectra at the
rest-frame wavelength of $\lambda_{\rm rest}$ $<$ 1215.67~\AA\ using a
principal component analysis (PCA) of low-redshift quasar spectra,
following the same manner as \citet{ish20}.  We fit the quasar's
spectra from the peak of \lya\ emission lines up to $\lambda_{\rm
  rest}$ $\sim$ 1600\AA, 2000\AA\ (or upper limits of the observed
spectra) using 7 -- 10 principal component spectra (PCS) of
\citet{suz05} or \citet{par11}.  In Figure~\ref{fig:norm}, we show the
result of PCA fits to the b/g quasar of PQ4 (MJD:55587) along with the
observed spectra as an example.  The PCA fits moderately reproduce a
global pattern of the continuum level in \lya\ forest, but they
over/under-estimate about half of the observed spectra with the
difference of $\lesssim$10\%.  Therefore, we manually fit a continuum
level to the PCA-fit normalized spectra again, using a low-order
spline function as done in \citet{pro13} and \citet{jal19}. Using the
final spectra, we measured the mean flux levels of \lya\ forest within
$\pm$10000~\kms\ of the $z_{\rm f/g}$, and confirmed that the
difference between the expected (from \citealt{fau08}) and the
observed flux levels distribute from $-$4.9\% to $+$7.8\% with the
average of $\sim$1\%\footnote{For PQ10 and PQ11, we masked spectral
  regions around DLA systems before measuring the flux level.}.
\citet{jal19} noticed that the manually fitted continuum level tends
to be under-estimated compared to the true level, especially for
low-resolution spectra of quasars at higher redshift ($z$ $>$ 2.6).
However, the under-estimation is almost negligible for the spectra of
our targets at relatively lower redshift ($\langle z \rangle$ $\sim$
2.3) with higher S/N ratio ($\langle$S/N$\rangle$ $\sim$
13~pixel$^{-1}$) since the level of under-estimation has a positive
(or negative) correlation with S/N ratio (or redshift), respectively
\citep{fau08}.  For quasars that have been observed more than once
(PQ1, PQ3, PQ9, and PQ11), we apply the normalization process above
for each exposure respectively and combine them to increase an S/N
ratio.  For PQ9, PQ10, and PQ11, we will use FOCAS and SDSS spectra
separately without combining them together to avoid any possible
systematic biases due to the difference in observing configurations.
Normalized spectra of 12 SDSS and 3 FOCAS spectra are presented in
Figures~\ref{fig:vel_plot} and \ref{fig:vel_plot2}.

%%% Figure 2 %%%

%%% Figure 3 %%%
\begin{figure*}[ht!]
  \begin{center}
    \includegraphics[width=8.5cm,angle=0]{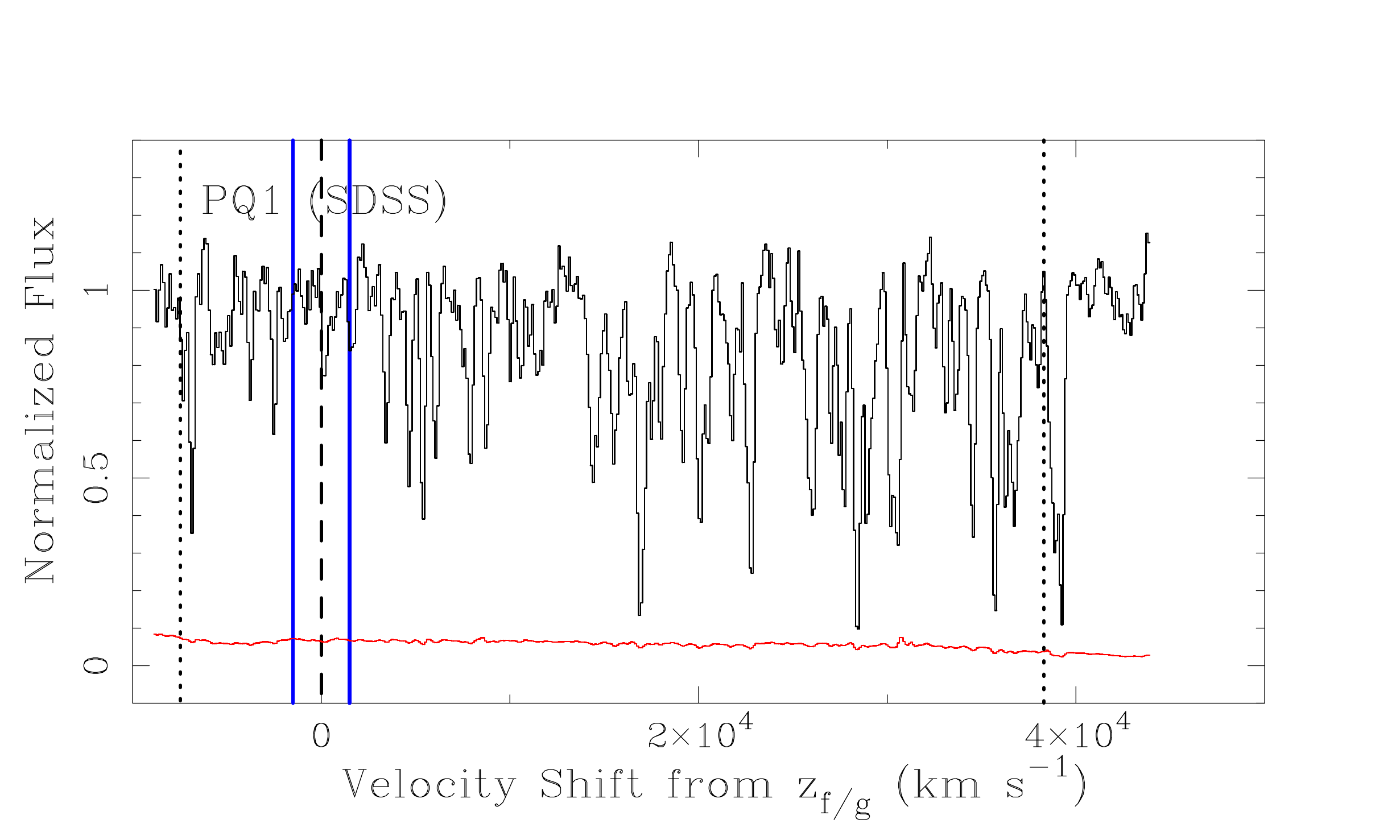} 
    \includegraphics[width=8.5cm,angle=0]{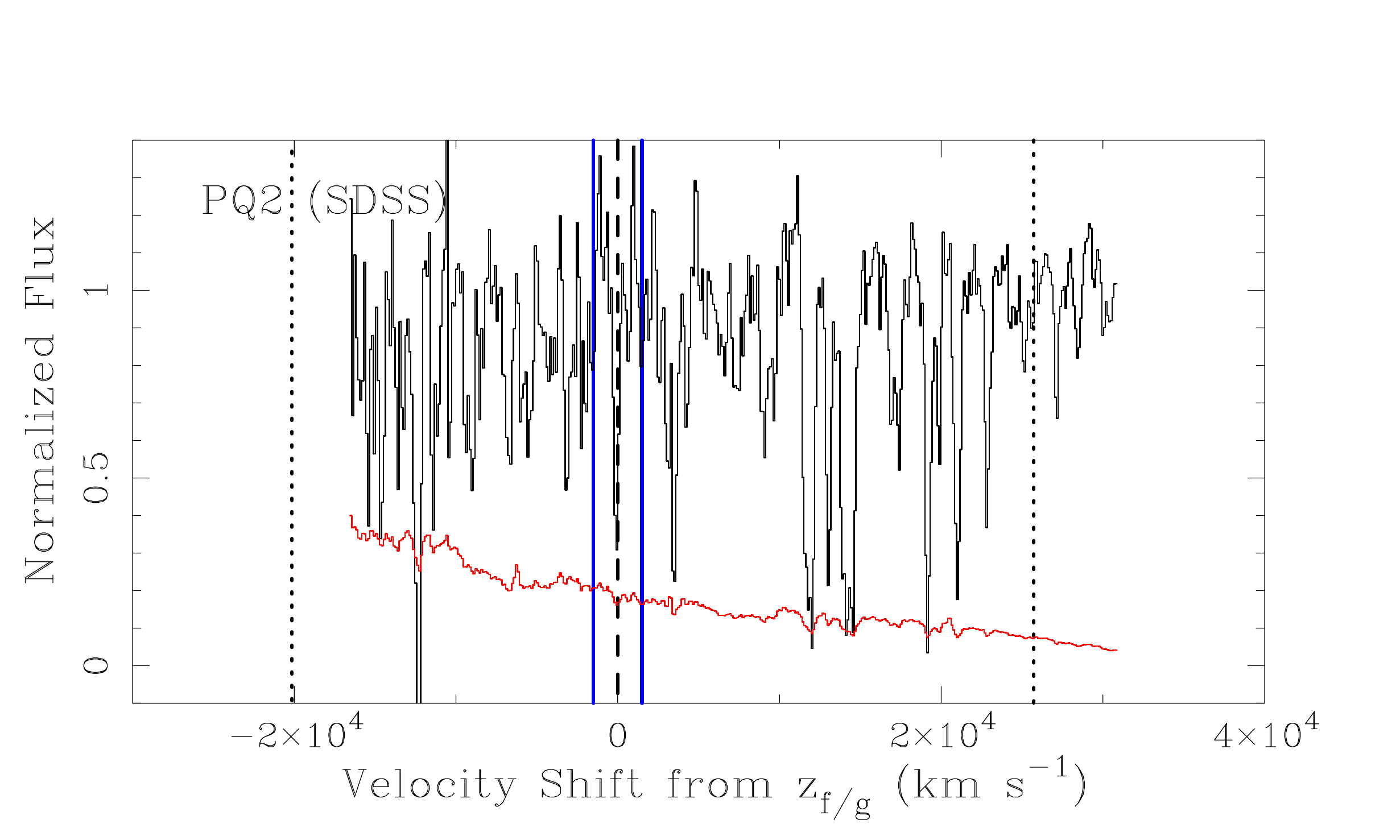}
    \includegraphics[width=8.5cm,angle=0]{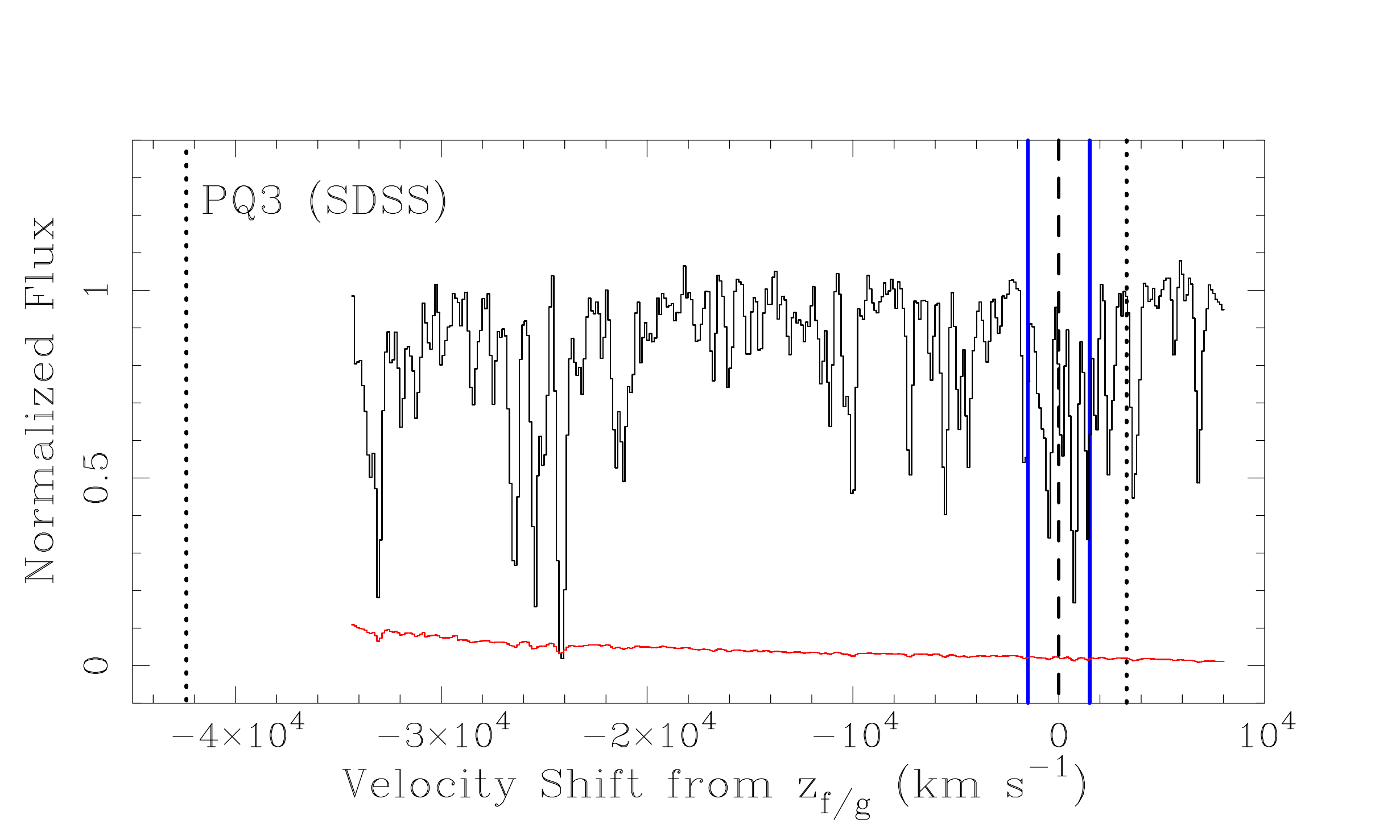}
    \includegraphics[width=8.5cm,angle=0]{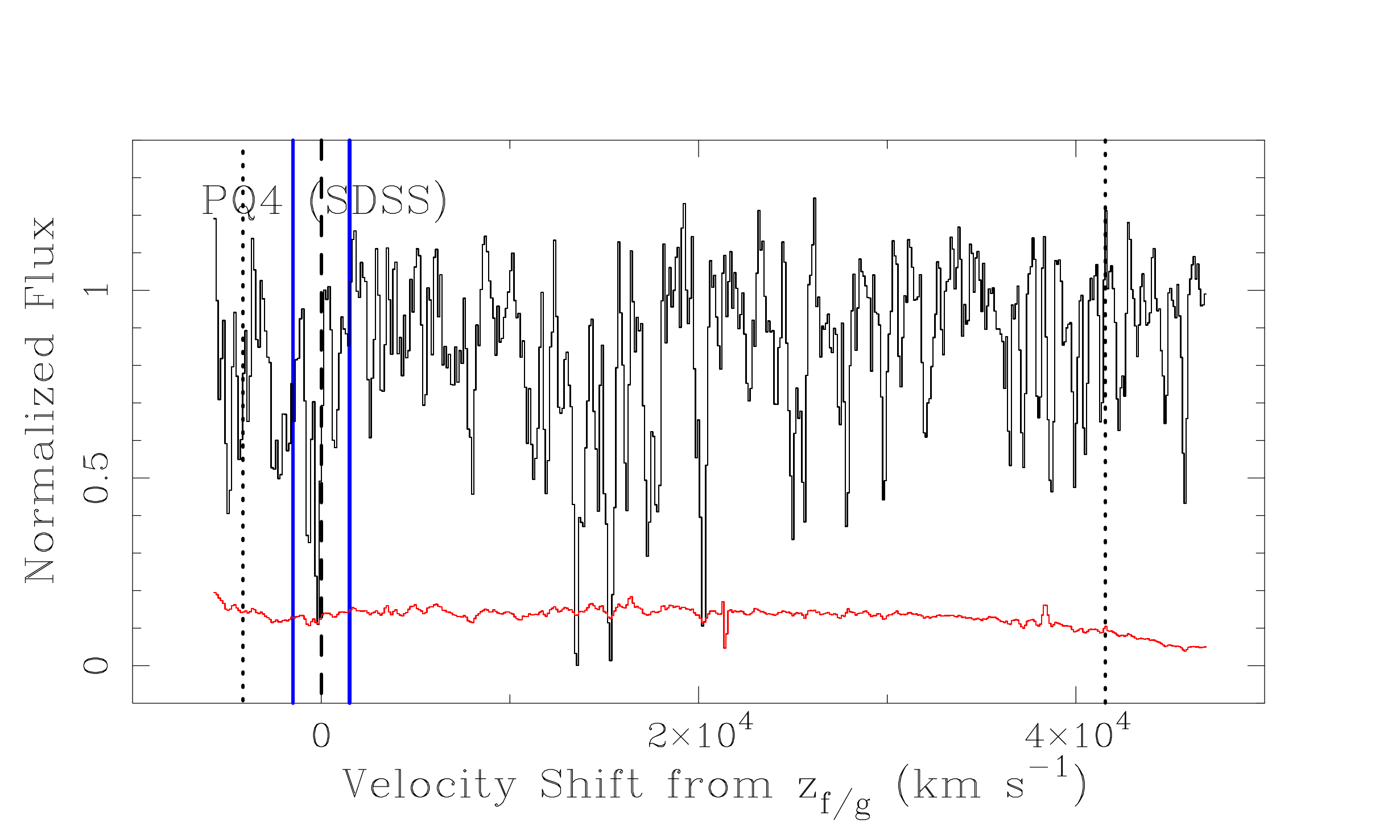}
    \includegraphics[width=8.5cm,angle=0]{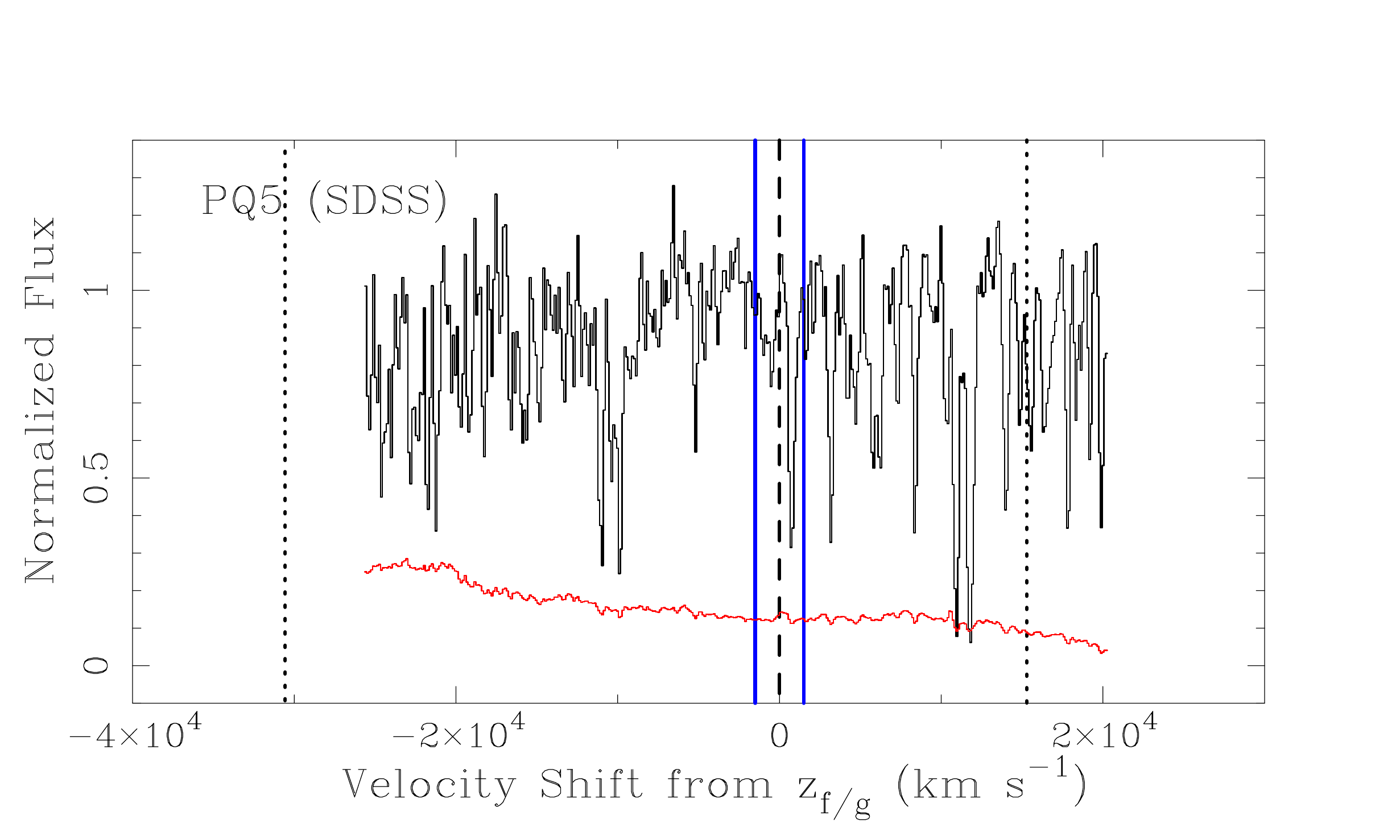}
    \includegraphics[width=8.5cm,angle=0]{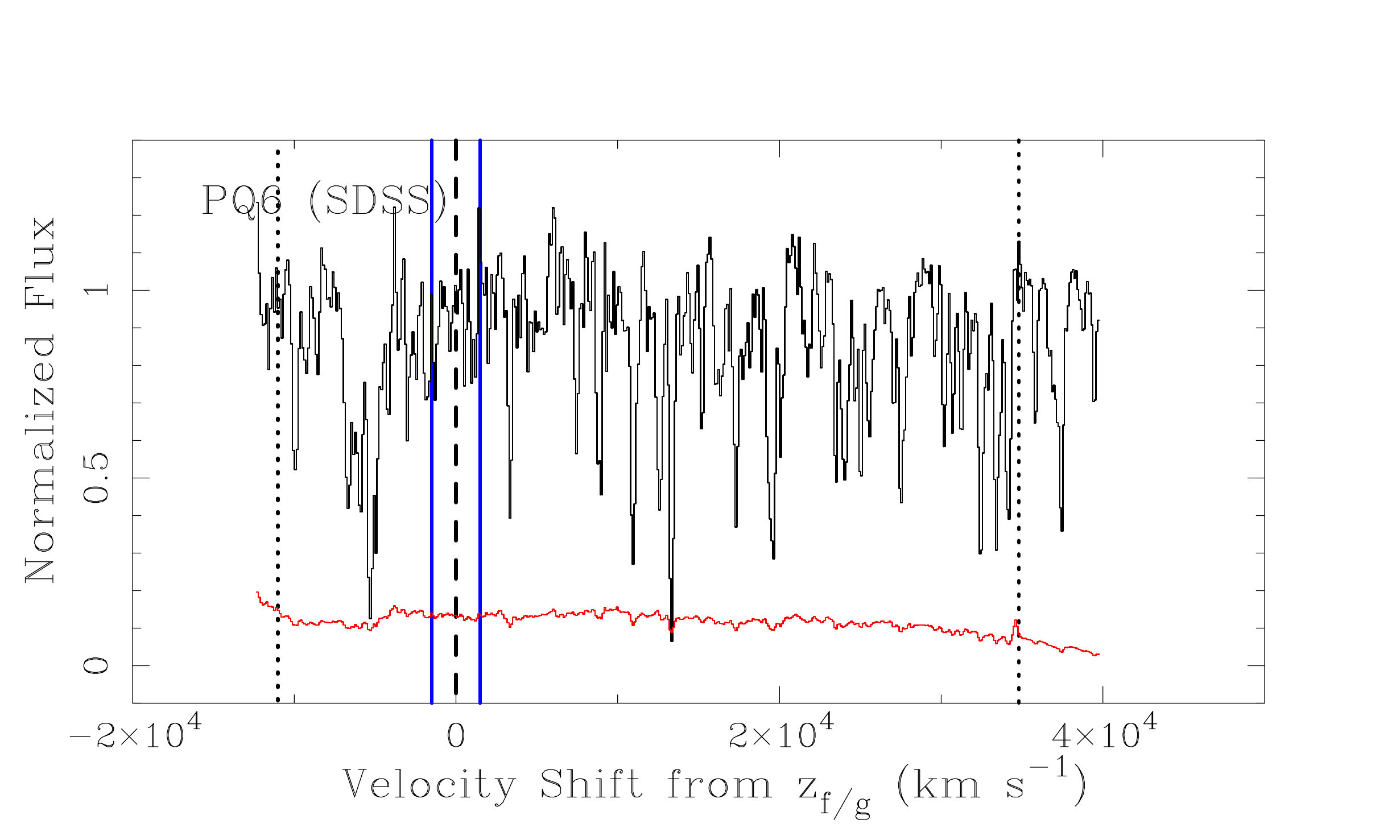}
    \includegraphics[width=8.5cm,angle=0]{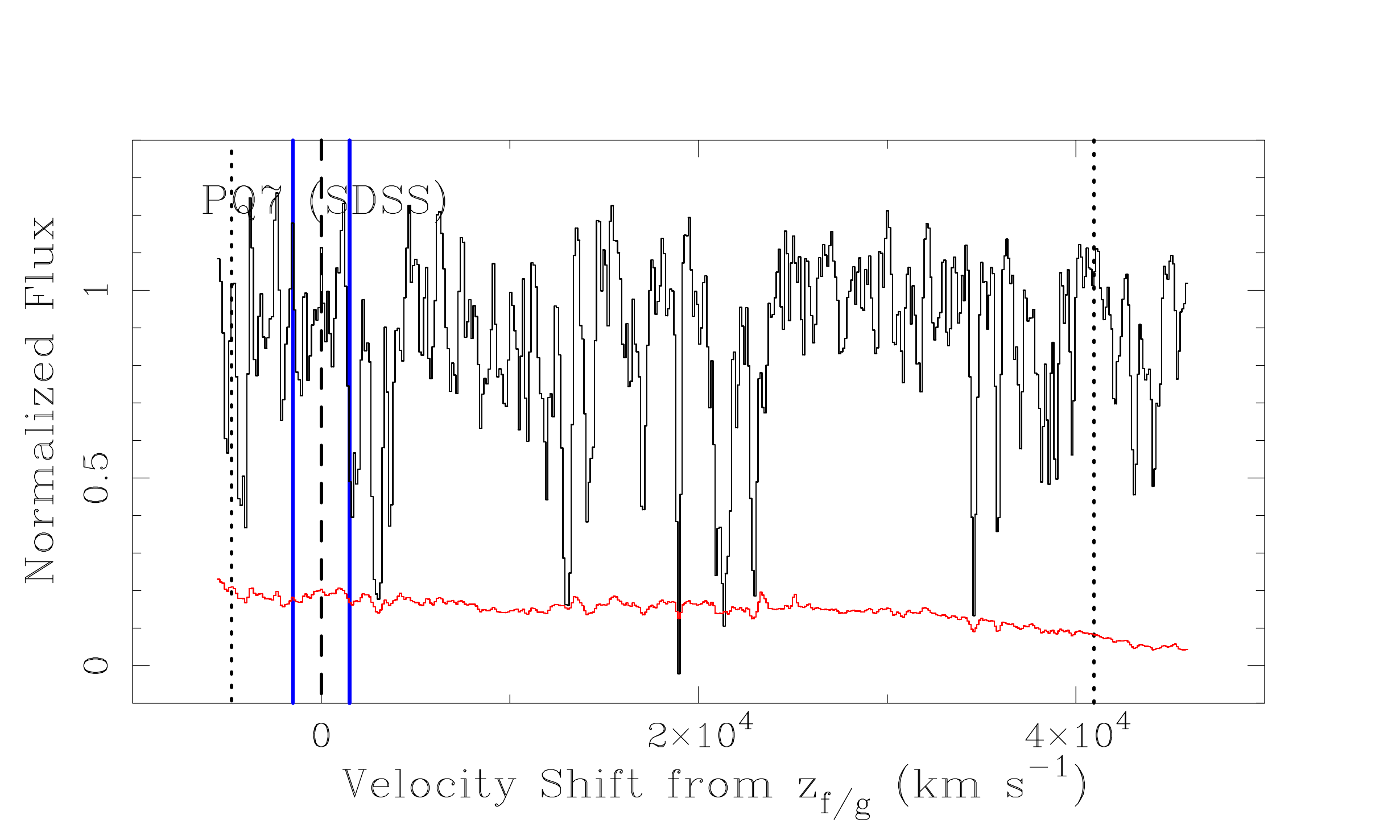}
    \includegraphics[width=8.5cm,angle=0]{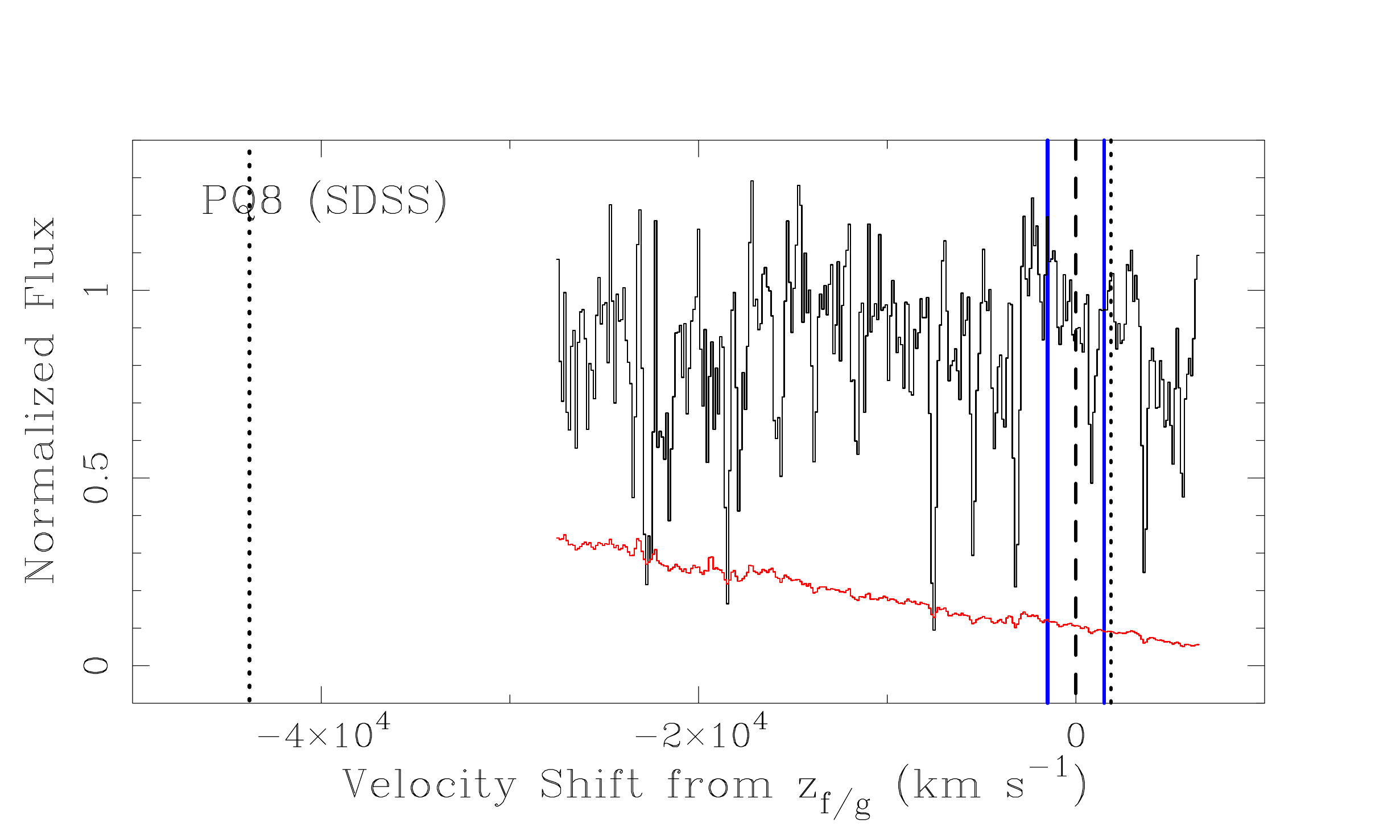}
  \end{center}  
  \caption{Normalized spectrum as a function of relative velocity from
    the redshift of f/g quasars as shown with a dashed vertical line.
    Blue solid vertical lines are $\pm 1500$~\kms\ from $z_{\rm
      f/g}$. Dotted vertical lines are the lower and upper limits of
    the optical depth analysis that correspond to \lyb\ emission
    (left) and 5000~\kms\ from $z_{\rm b/g}$
    (right).\label{fig:vel_plot}}
  \vspace{1cm}
\end{figure*}

\begin{figure*}[ht!]
\setcounter{figure}{2}
  \begin{center}
    \includegraphics[width=8.5cm,angle=0]{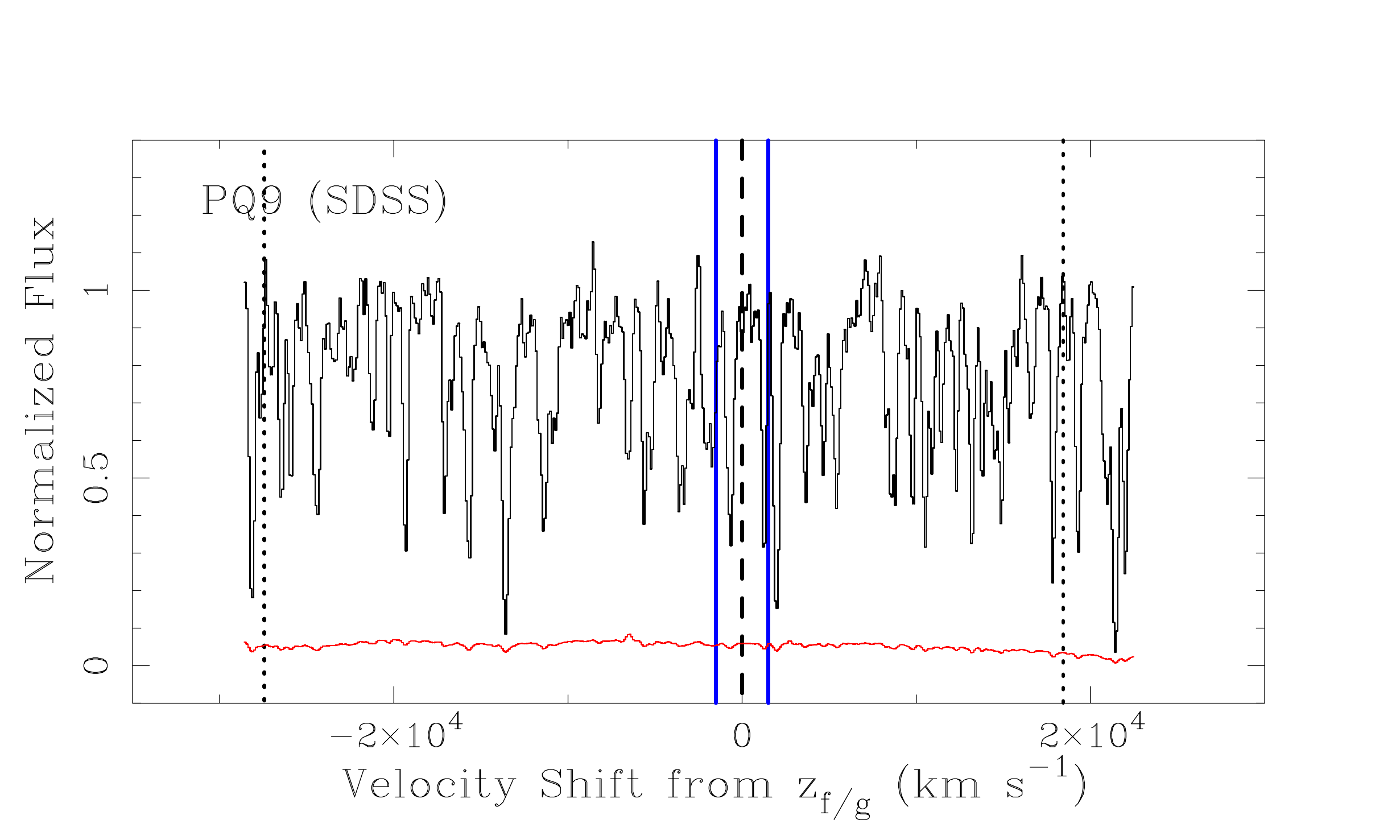}
    \includegraphics[width=8.5cm,angle=0]{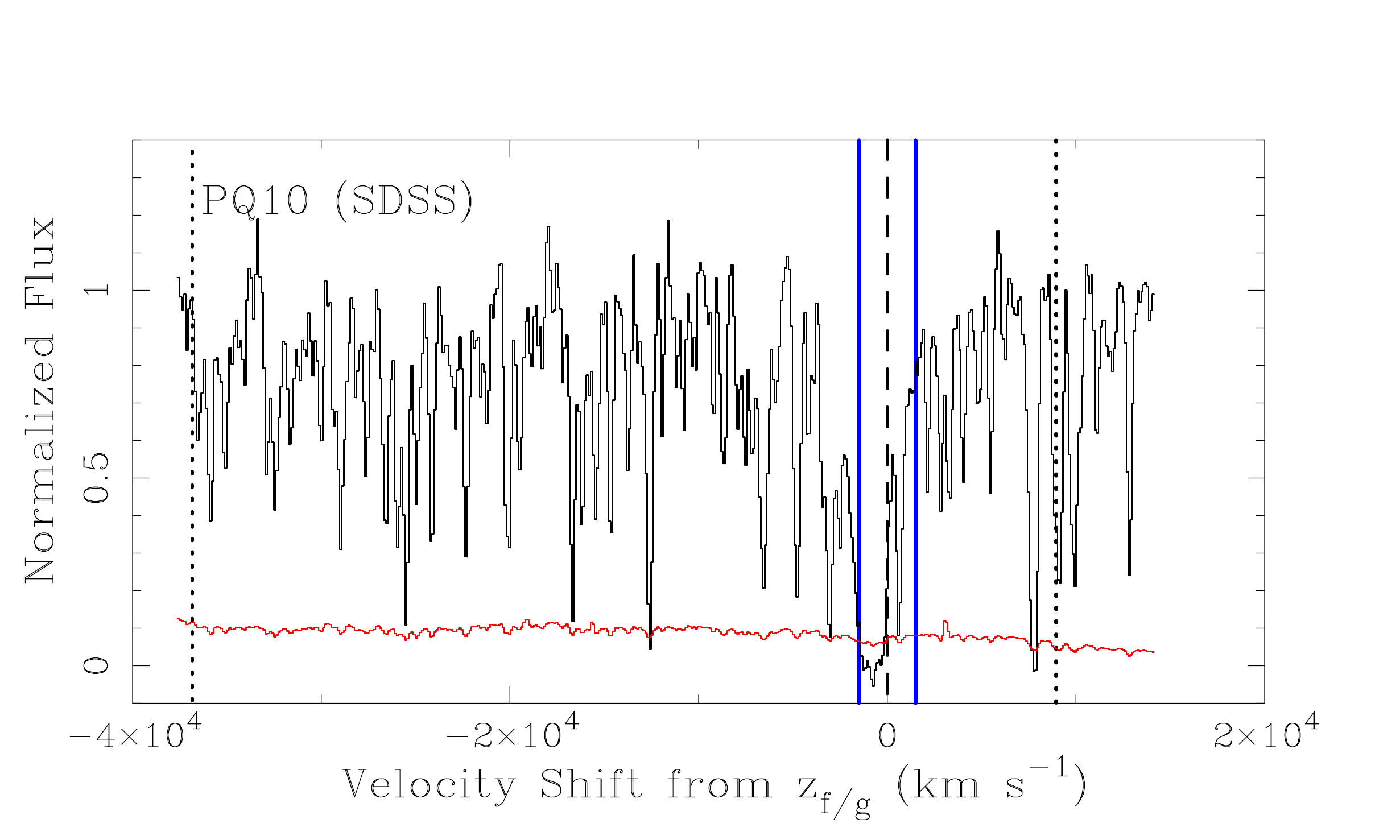}
    \includegraphics[width=8.5cm,angle=0]{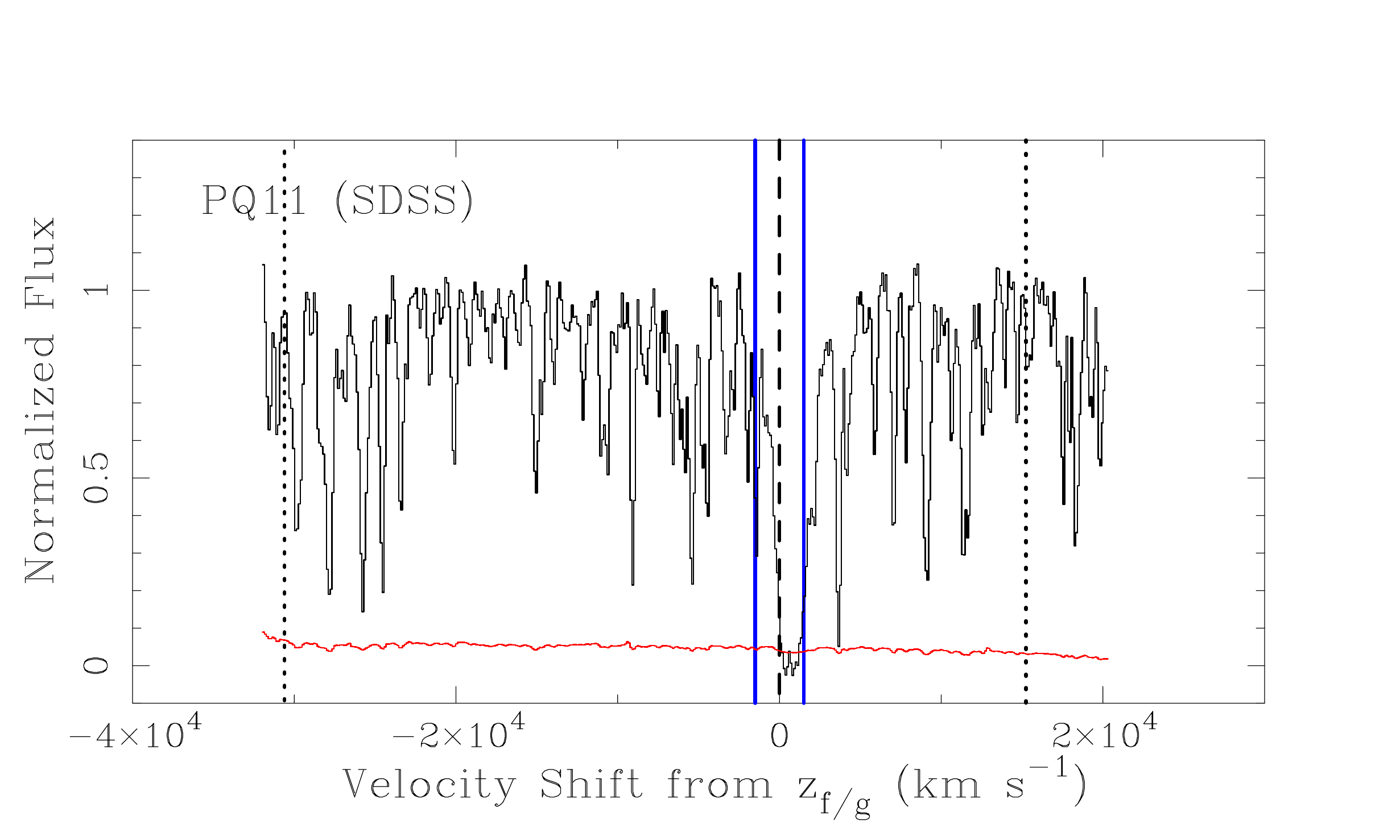}
    \includegraphics[width=8.5cm,angle=0]{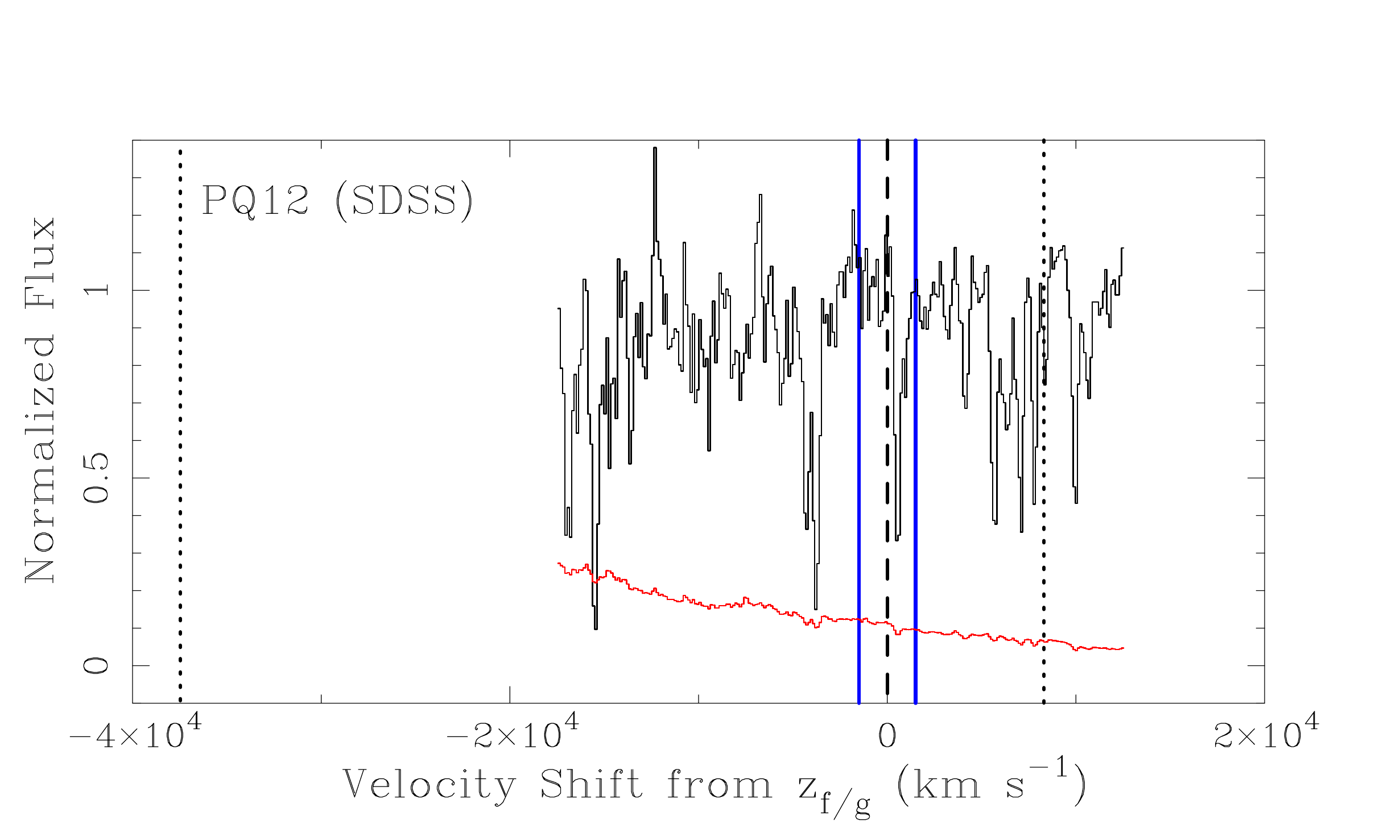}
  \end{center}  
  \caption{Continued.}
\end{figure*}

%%% Figure 4 %%%
\begin{figure*}[ht!]
  \begin{center}
    \includegraphics[width=8.5cm,angle=0]{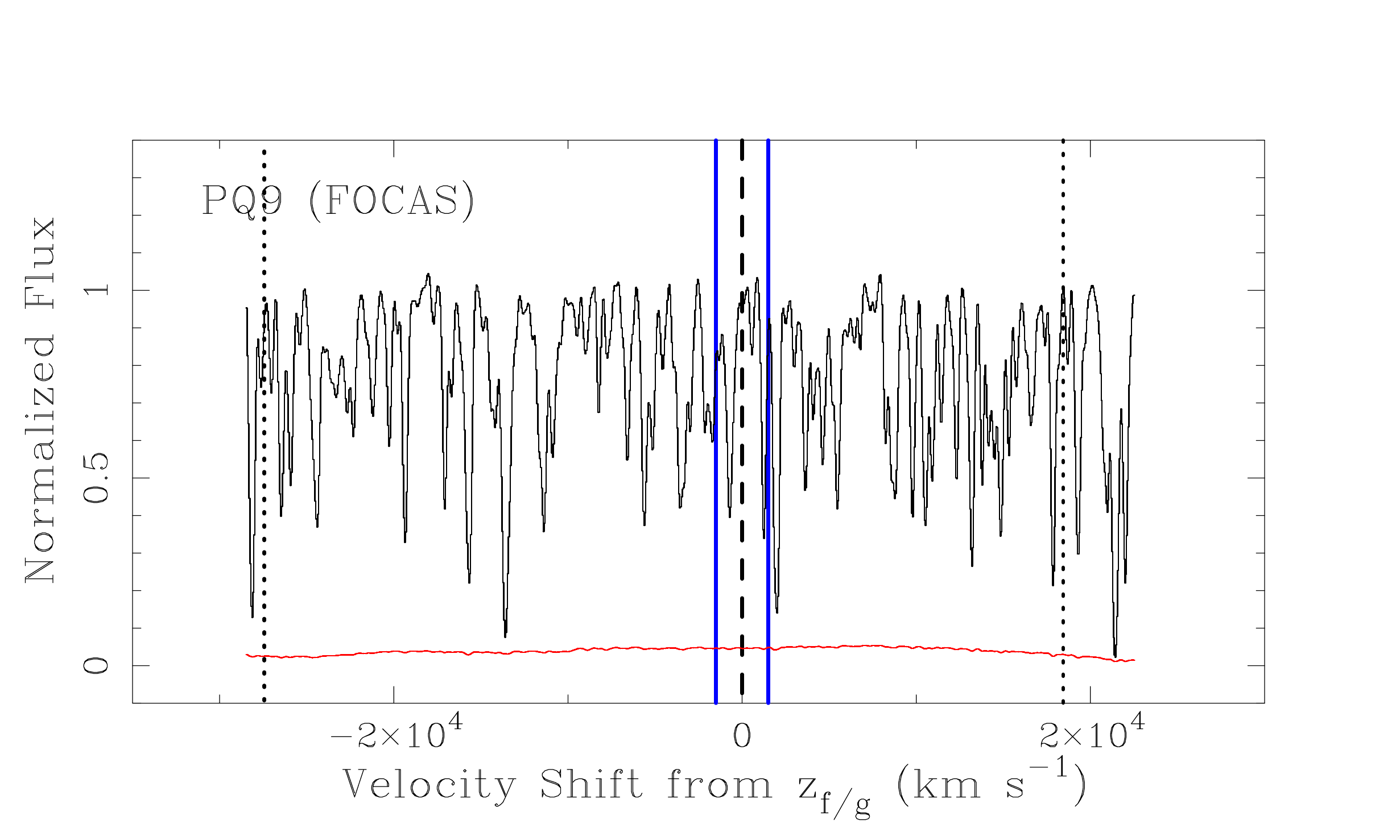}
    \includegraphics[width=8.5cm,angle=0]{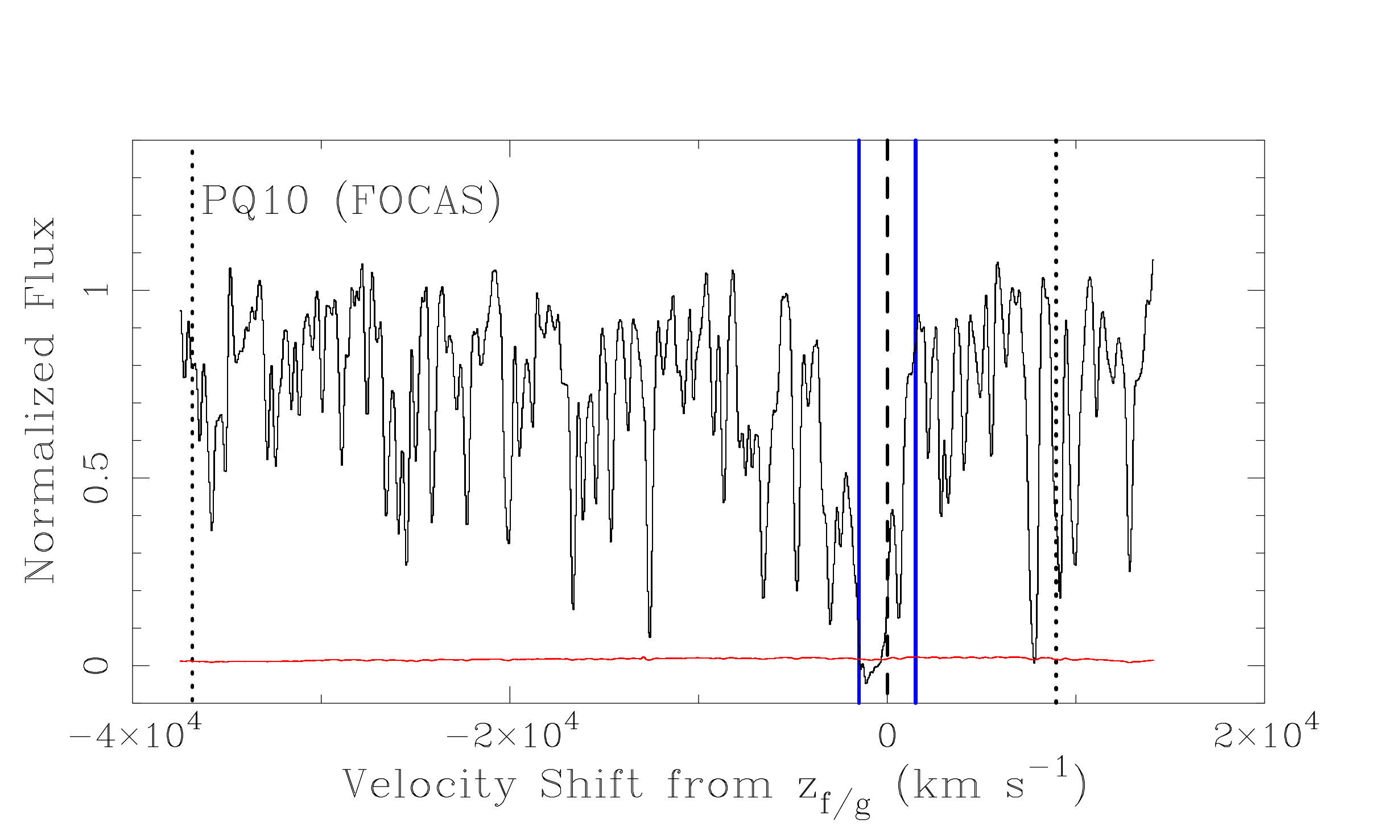}
    \includegraphics[width=8.5cm,angle=0]{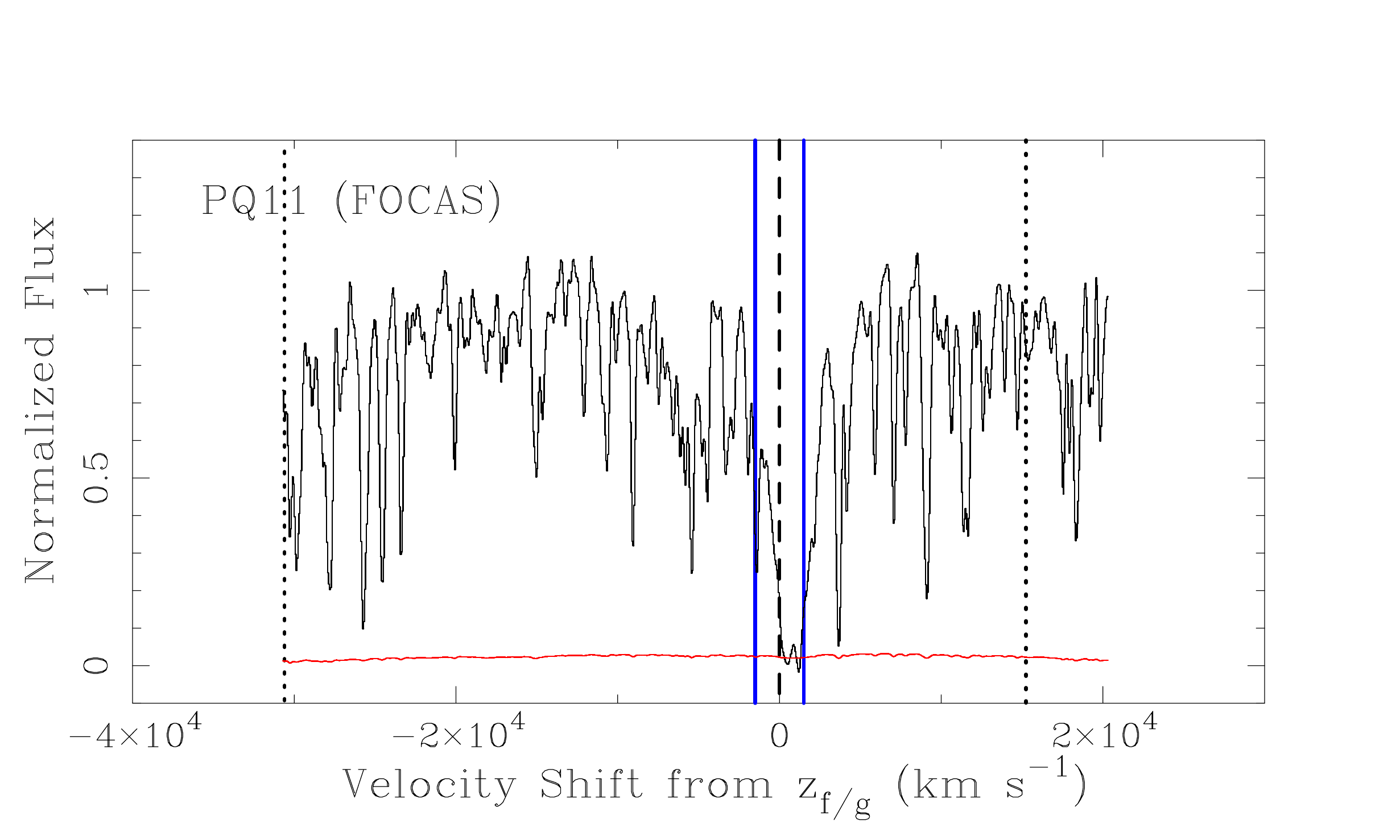}
  \end{center}  
  \caption{Same as Figure~\ref{fig:vel_plot}, but for PQ9, PQ10, and
    PQ11 spectra taken with Subaru/FOCAS.\label{fig:vel_plot2}}
  \vspace{1cm}
\end{figure*}

\subsection{Systemic Redshift of foreground quasars} \label{sec:zsys}
We need accurate estimation of the systemic redshift of foreground
quasars for the analyses of TPE. \citet{lyk20} presented a primary
redshift as the best estimation of systemic redshift among several
methods including visual inspection, pipeline, neural network, and so
on.  The primary redshift of all quasars (except for the f/g quasar of
PQ4\footnote{Redshift of this quasar was estimated using a pipeline in
  \citet{lyk20}, but we confirm the measured redshift is consistent to
  the emission redshift of \ion{Mg}{2}.}) are visually inspected based
on the peak of the \ion{Mg}{2} emission lines \citep{par17}. The
\ion{Mg}{2}-based redshifts are on average 57~\kms\ blueshifted from
systemic redshifts based on stellar \ion{Ca}{2} absorption lines
\citep{she16} and their total uncertainty is $\sim$300~\kms\ (i.e.,
$\delta z$ $\sim$ 0.003 at $z$ $\sim$ 2) including intrinsic,
systematic, and statistical errors \citep{she16,jal19}. We adopted the
same value as the typical uncertainty of systemic redshift as shown in
Table~\ref{tab:qsos}.  Since the velocity shift of the
\ion{Mg}{2}-based redshift from the systemic redshift is small enough
and smaller than its own uncertainty by a factor of $\sim$5, we regard
the primary redshift in \citet{lyk20} as the systemic redshift in this
study without any corrections.

\subsection{Enhancement of quasar ionizing photon flux} \label{sec:enhance}
An ionization condition of CGM and IGM around the f/g quasars depends
on the total number of hydrogen-ionizing photons emitted by the f/g
quasars and a distance from them in addition to the extra galactic
ionizing background. We calculate the Lyman continuum luminosity of
the f/g quasar at $\lambda_{\rm rest}$ = 912\AA\ in the quasar's rest
frame assuming isotropic radiation by

\begin{equation}
L_{\rm \nu}^{912} = F_{\rm \nu}^{912} \times 4 \pi D_L^2,
\end{equation}
where $F_{\nu}^{912}$ is the specific flux at $\lambda_{\rm rest}$ =
912\AA\ and $D_L$ is the luminosity distance of the f/g quasar, using
the same prescription as described in \citet{jal19}. The value of
$F_{\nu}^{912}$ is estimated from the observed flux at $\lambda_{\rm
  obs}$ $\sim$ 1325$\times$(1+$z_{\rm f/g}$)~\AA, and extrapolated to
$\lambda_{\rm rest}$ = 912~\AA\ by using a broken power-law
\citep{kha15} with a power-law index of $\alpha$ = 0.44 \citep{van01}
at $\lambda_{\rm rest}$ $>$ 1300~\AA\ and 1.57 \citep{tel02} at
$\lambda_{\rm rest}$ $<$ 1300~\AA.  If we instead use the latest
quasar spectral energy distribution (SED) with $\alpha$ = 0.61
\citep{lus15} at $\lambda_{\rm rest}$ $>$ 912~\AA, $F_{\nu}^{912}$ and
$L_{\rm \nu}^{912}$ become larger by a factor of $\sim$1.4. Throughout
the paper, we use the original SED above matching to those used in
\citet{pro13} and \citet{jal19}.

\begin{figure}[ht!]
  \begin{center}
    \includegraphics[width=8cm,angle=0]{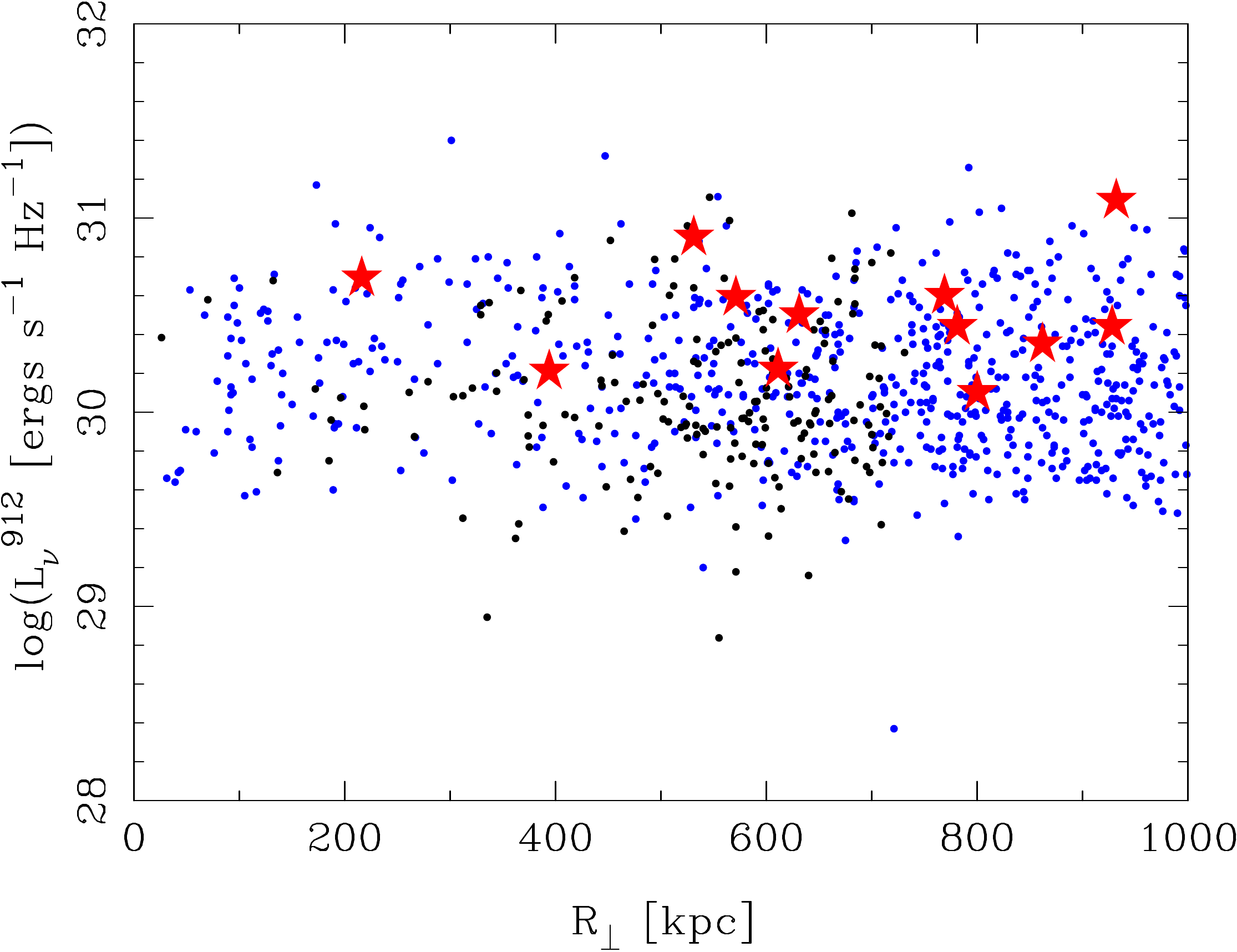}
  \end{center}  
  \caption{Scatter plot of the Lyman continuum luminosity of the f/g
    quasars at $\lambda_{\rm rest}$ = 912\AA\ ($L_{\nu}^{912}$) and
    the perpendicular distance between the f/g quasar and the
    sight-line toward the b/g quasar ($R_{\perp}$).  Red filled stars
    are BAL quasars in our sample, while blue and black dots are
    non-BAL quasars from \citet{pro13} and \citet{jal19},
    respectively.\label{fig:dist}}
\end{figure}

We also calculate the shortest distance between the f/g quasar and the
sight-line toward the b/g quasar by

\begin{equation}
R_{\perp} = D_A \times \theta,
\end{equation}
where $D_A$ is the angular diameter distance of the f/g quasar and
$\theta$ is the observed angular separation between the f/g and b/g
quasars.  We confirm that our sample is not biased, since the
distributions of $L_{\rm \nu}^{912}$ and $R_{\perp}$ of our BAL
quasars are similar to those of non-BAL quasars used in \citet{pro13}
and \citet{jal19}, as shown in Figure~\ref{fig:dist},

Using $L_{\rm \nu}^{912}$ and $R_{\perp}$ above, we calculate the
enhancement of the quasar ionizing photon flux ($F_{\rm qso}$) over
that of the extragalactic ionizing background ($F_{\rm uvb}$) at a
distance of $R_{\perp}$ by,

\begin{equation}
g_{\rm UV} = \frac{F_{\rm qso} + F_{\rm uvb}}{F_{\rm uvb}},
\end{equation}
assuming the quasar emission is isotropic. \citet{hen06} calculated
$g_{\rm UV}$ assuming a power-law index of $\alpha$ = 1.8 in $J_{\rm
  \nu}$ $\propto$ $\nu^{-\alpha}$ for the mean specific intensity of
the extragalactic ionizing background. In this study, we also
calculate the enhancement of the ionizing flux using the spectral
energy distribution (SED) of $J_{\rm \nu}$ as computed by
\citet{har12} ($g_{\rm UV}$$^{\prime}$, hereafter) in addition to the
original one, $g_{\rm UV}$.

Here, we should note that the photoionization {\it rate} depends not
only on the ionizing photon flux, but on the \ion{H}{1}
photoionization cross section that is inversely proportional to the
cube of frequency (i.e., $\sigma_{\rm HI}(\nu)$ $\propto$
$\nu^{-3}$). Therefore, we calculate the \ion{H}{1} ionization rates
by the quasar's ionizing photon flux ($\Gamma_{\rm qso}$) and that due
to the extragalactic background ionizing photon flux ($\Gamma_{\rm
  uvb}$). The enhancement of the $\Gamma_{\rm qso}$ over the
$\Gamma_{\rm uvb}$ at a distance of $R_{\perp}$ is defined as
    
\begin{equation}
1 + \omega_{\rm r} = \frac{\Gamma_{\rm uvb} + \Gamma_{\rm
    qso}}{\Gamma_{\rm uvb}}.
\end{equation}

The values of $g_{\rm UV}$ and $g_{\rm UV}$$^{\prime}$ are slightly
different from each other, depending on the SED of the extragalactic
background ionizing flux that we assume.  On the other hand, the value
of $1 + \omega_{\rm r}$ is almost independent of the SED. So, we will
use $1 + \omega_{\rm r}$ (instead of $g_{\rm UV}$) for discussing the
influence from the quasar's ionizing radiation.  The values of
parameters we calculated are summarized in Table~\ref{tab:pe}.

%%% Figure 5 %%%

%%% Table 3 %%%
\begin{deluxetable*}{ccccccccc}
\tablecaption{Parameters of Transverse Proximity Effect \label{tab:pe}}
\tablewidth{0pt}
\tablehead{
\colhead{ID}                       &
\colhead{$\log L_{\rm \nu}^{912}$(f/g)$^a$} &
\colhead{$R_{\perp}^b$}              &
\colhead{$g_{\rm UV}$$^c$}           &
\colhead{$g_{\rm UV}^{\prime}$$^d$}    &
\colhead{($1+\omega_{\rm r}$)$^e$}   &
\colhead{$\Delta v^f$}             &
\colhead{$\log N_{\rm HI}$$^g$}      &
\colhead{DLA$^h$}                  \\
\colhead{}                         &
\colhead{(erg/s)}                  &
\colhead{(kpc)}                    &
\colhead{}                         &
\colhead{}                         &
\colhead{}                         &
\colhead{(\kms)}                   &
\colhead{(\cmm)}                   &
\colhead{}                         \\
}
\startdata
PQ1  & 30.90 & 531 & 263.8 &  61.3 &  54.8 &     190          & 13.88$^{+0.11}_{-0.12}$ & \\
PQ2  & 30.69 & 216 & 995.1 & 228.7 & 205.0 &   $-$86          & 14.43$^{+0.13}_{-0.14}$ & \\	
PQ3  & 30.35 & 862 &  29.0 &   7.4 &   6.7 &     750          & 14.61$^{+0.01}_{-0.01}$ & \\
PQ4  & 30.22 & 611 &  43.0 &  10.6 &   9.6 &  $-$168          & 14.63$^{+0.12}_{-0.11}$ & \\
PQ5  & 30.21 & 394 &  99.5 &  23.6 &  21.2 &     747          & 14.52$^{+0.07}_{-0.07}$ & \\
PQ6  & 30.44 & 781 &  44.0 &  10.9 &   9.8 & $-$1296          & 13.84$^{+0.24}_{-0.43}$ & \\
PQ7  & 30.50 & 631 &  75.6 &  18.1 &  16.3 & $-$1111          & 13.89$^{+0.30}_{-0.78}$ & \\
PQ8  & 31.09 & 932 & 133.2 &  31.4 &  28.1 &     835          & 14.26$^{+0.09}_{-0.09}$ & \\
PQ9  & 30.59 & 571 & 120.4 &  28.7 &  25.2 &    1269   (1249) & 14.35$^{+0.02}_{-0.02}$ & \\
PQ10 & 30.60 & 769 &  66.0 &  16.0 &  14.2 &  $-$829 ($-$839) & 20.62$^{+0.01}_{-0.01}$ & $z_{\rm DLA}$ = 2.539 \\
PQ11 & 30.44 & 928 &  31.4 &   8.0 &   7.2 &     746    (768) & 20.75$^{+0.00}_{-0.00}$ & $z_{\rm DLA}$ = 2.443 \\
PQ12 & 30.10 & 800 &  19.3 &   5.2 &   4.7 &     561          & 14.44$^{+0.06}_{-0.06}$ 
\enddata
\tablenotetext{a}{Specific luminosity at $\lambda_{\rm rest}$ =
  912\AA.}
\tablenotetext{b}{Shortest distance between the f/g quasar and the
  sightline toward the b/g quasar.}
\tablenotetext{c}{Enhancement of the quasar's ionizing photon flux
  over that of the extragalactic ionizing background at a distance of
  $R_{\perp}$ from foreground quasar, assuming $J_{\rm \nu} \propto
  \nu^{-1.8}$ as an SED of the extragalactic ionizing background
  \citep{hen06}.}
\tablenotetext{d}{Same as $g_{\rm UV}$, but using the version of
  $J_{\rm \nu}$ computed by \citet{har12} for the extragalactic
  ionizing background.}
\tablenotetext{e}{Enhancement of the \ion{H}{1} ionization rate
  including the quasar's ionizing photon over that due to only the
  extragalactic ionizing background at a distance of $R_{\perp}$ from
  a foreground quasar, using the version of $J_{\rm \nu}$ computed by
  \citet{har12} for the extragalactic ionizing background.}
\tablenotetext{f}{Relative velocity between $z_{\rm f/g}$ and $z_{\rm
    abs}$ of the strongest \ion{H}{1} line in $\pm$1500~\kms\ of
  $z_{\rm f/g}$. Numbers in parentheses are measured in FOCAS
  spectra.}
\tablenotetext{g}{\ion{H}{1} column density of the strongest
  absorption line in $\pm$1500~\kms\ of $z_{\rm f/g}$, which are
  measured in FOCAS spectra for PQ9, PQ10, and PQ11 and SDSS spectra
  for the other quasars.}
\tablenotetext{h}{Redshift of Damped \lya\ system detected in
  $\pm$1500~\kms\ of $z_{\rm f/g}$, as discovered by \citet{lyk20}.}
\end{deluxetable*}

\section{Results} \label{sec:results}
In this section, we examine the \ion{H}{1} absorption strength in
transverse direction through three analyses following \citet{pro13}:
i) measuring the detection rate (i.e., covering fraction, $C_{\rm f}$)
of optically thick absorbers around BAL quasars, ii) studying how the
\ion{H}{1} column density at $z_{\rm abs}$ $\sim$ $z_{\rm f/g}$
correlates with other parameters, iii) examining the \ion{H}{1}
absorption strength at $z_{\rm abs}$ $\sim$ $z_{\rm f/g}$ in the
stacked spectrum of the b/g quasars.

\subsection{Detection Rate of Optically Thick \ion{H}{1} Absorbers}\label{sec:rate}
Since systemic redshifts that are measured based on \ion{Mg}{2}
emission lines have their own typical uncertainties with $\delta v$ of
$\sim$300~\kms, the positions of \ion{H}{1} absorption lines
corresponding to the CGM and IGM around f/g quasars can be shifted
accordingly.  The halo gas of f/g quasar also has it own peculiar
motion.  Therefore, we assume the {\it strongest} absorption line in
$\pm$1500~\kms\ of $z_{\rm f/g}$ as the systems corresponding to the
CGM/IGM around f/g quasars following \citet{pro13}.  We also measure
their column densities using the Voigt profile fitter {\tt mc2fit}
\citep{ish21} with absorption redshift, column density, and Doppler
parameter as free parameters.  The relative velocity of the strongest
\ion{H}{1} line from $z_{\rm f/g}$ and their column density are
summarized in Table~\ref{tab:pe}.

\begin{figure}[ht!]
  \begin{center}
    \includegraphics[width=8cm,angle=0]{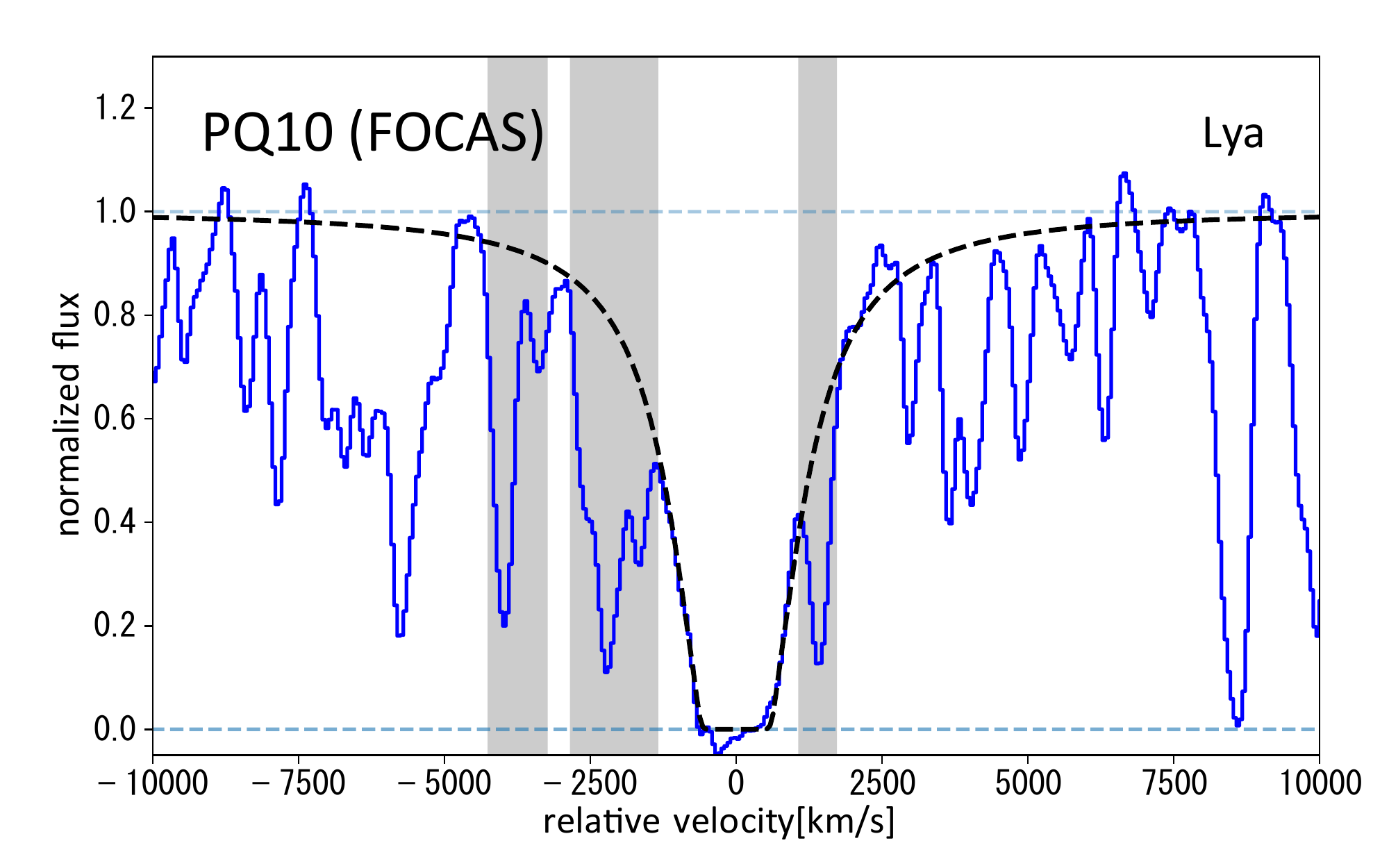}
    \includegraphics[width=8cm,angle=0]{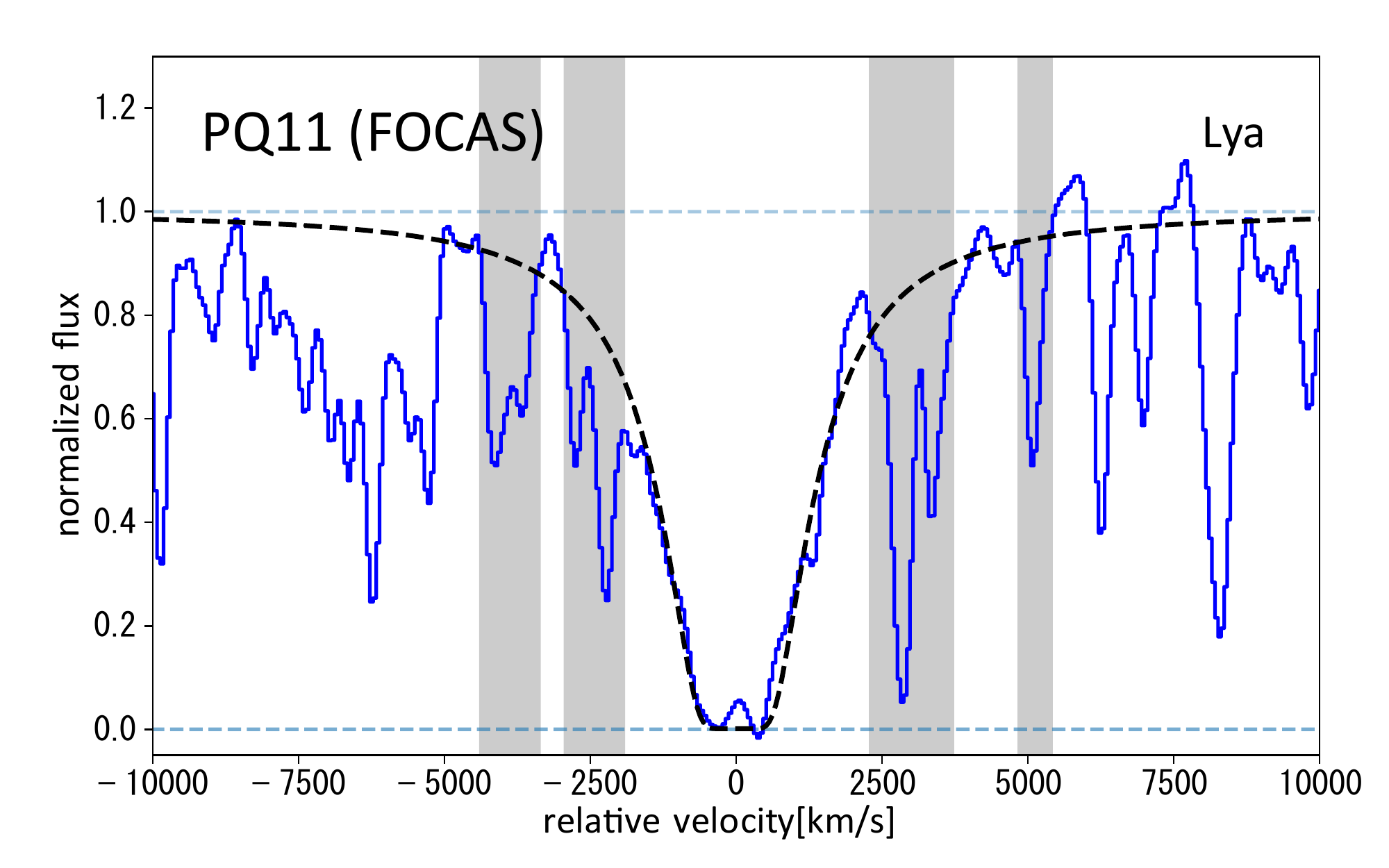}
  \end{center}  
  \caption{Observed spectrum (blue histogram) and best fit model
    (black dashed curve) of strong \ion{H}{1} absorption lines at
    $z_{\rm abs}$ $\sim$ $z_{\rm f/g}$ in b/g quasars of PQ10 (top)
    and PQ11 (bottom).  Shaded areas are ignored while model fitting
    due to blending with other lines.\label{fig:dla}}
\end{figure}

As already reported in \citet{lyk20}, we confirmed there exist
candidates of DLA systems with column densities of $\log N_{\rm
  HI}$/(cm$^{-2}$) $\gtrsim$ 20 in spectra of b/g quasars of PQ10 and
PQ11 as shown in Figure~\ref{fig:dla}.  Although these systems show
damping wings, they could be modeled with multiple weaker components
since their line profiles are not continuously saturated at line
centers, especially for the system in PQ11.  Indeed, a line blending
in low-resolution spectrum tends to over-estimate the column density.
We could not examine the existence of Lyman limits either (i.e., a
reliable indicator of DLA systems) since they are outside of our SDSS
and FOCAS spectra.  However, we detected several low-ionized metal
absorption lines (e.g., \ion{O}{1}1302, \ion{C}{2}1335,
\ion{Si}{2}1260, \ion{Si}{2}1527, and \ion{Si}{4}1394,1403) at the
redshift of these systems, which suggests that they are closely
related to galaxies.  Thus, we are not confident that our measured
values of $\log N_{\rm HI}$ are absolutely correct, but their large
equivalent widths of EW$_{\rm rest}$ $>$ 10\AA\ (which corresponds to
$\log N_{\rm HI}$/(cm$^{-2}$) $>$ 20 at the square-root part of the
curve of growth) and the existence of metal absorption lines suggest
that their optical depths are at least larger than those of the
optically thick absorbers around non-BAL quasars \citep{pro13} with
EW$_{\rm rest}$ $\sim$ 1--2\AA\ (which corresponds to $\log N_{\rm
  HI}$/(cm$^{-2}$) $\gtrsim$ 17 at the flat part of the curve of
growth).

Thus, the detection rate (i.e., the covering fraction) of optically
thick gas around BAL quasars is $C_{\rm f}$ = 2/12 $\sim$
$0.17^{+0.22}_{-0.11}$ at $R_{\perp}$ $\sim$ 220 -- 930~kpc, assuming
Poisson noise \citep{geh86}. Although our $C_{\rm f}$ estimate has a
large uncertainty due to the small sample size, the fraction is
consistent to the value estimated at similar transverse distance and
velocity interval from non-BAL quasars ($C_{\rm f}$ $\sim$
0.19$\pm$0.02; \citealt{pro13}). We will discuss a large difference in
the typical column density of \ion{H}{1} absorbers around BAL and
non-BAL quasars in Section~\ref{sec:anisotropy}.

%%% Figure 6 %%%

\subsection{Correlations among TPE parameters} \label{sec:correlations}
\begin{figure*}[ht!]
  \begin{center}
    \includegraphics[width=17cm,angle=0]{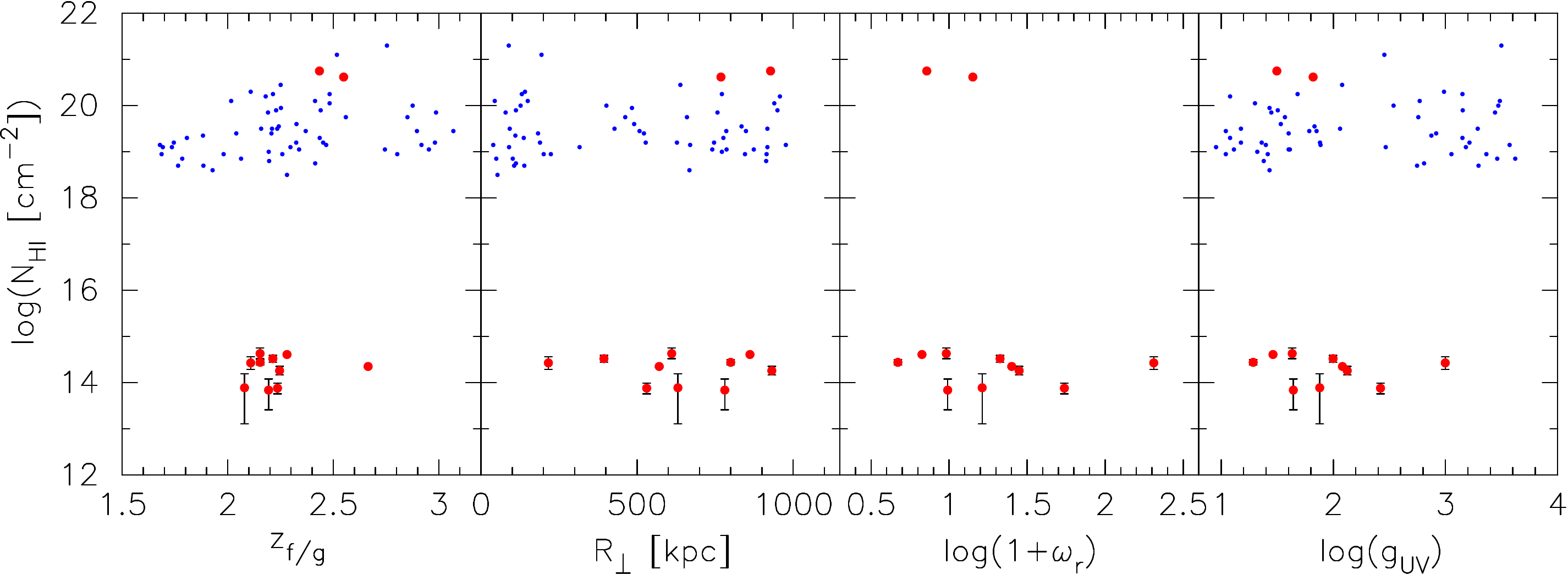}
  \end{center}  
  \caption{Column density distribution of the strongest \ion{H}{1}
    absorption line in $\pm$1500~\kms\ of $z_{\rm f/g}$ as functions
    of redshift of foreground quasar ($z_{\rm f/g}$), transverse
    distance between f/g quasars and lines of sight to b/g quasars
    ($R_{\perp}$), enhancement of the \ion{H}{1} ionization rate
    (1+$\omega_{\rm r}$) and the \ion{H}{1} ionization radiation flux
    ($g_{\rm UV}$) due to the quasar's ionizing photon over that due
    to only the extragalactic ionization background at $R_{\perp}$.
    Filled red circles denote BAL quasars (this study), while blue
    dots are optically thick absorbers around non-BAL quasars
    \citep{pro13}. Note that more than 80\%\ of quasars in
    \citet{pro13} are not plotted here, since only upper limits of
    $\log N_{\rm HI}$ are placed due to their low S/N
    spectra.\label{fig:correlation}}
\end{figure*}

Figure~\ref{fig:correlation} summarizes the column density of the
strongest \ion{H}{1} line in $\pm$1500~\kms\ of $z_{\rm f/g}$ as
functions of redshift of foreground quasar ($z_{\rm f/g}$), the
transverse distance ($R_{\perp}$), the enhancement of the ionization
rate due to quasar's ionizing photons (1+$\omega_{\rm r}$), and the
enhancement of the quasar ionizing flux radiation ($g_{\rm UV}$).  As
described above, only two b/g quasars (i.e., PQ10 and PQ11) among 12
have absorption systems arising in optically thick absorbers, while
the others have only weaker lines with column densities of $\log
N_{\rm HI}$/(cm$^{-2}$) $<$ 15. Thus, there is a large gap (about five
orders of magnitude) in the column densities of the strongest
\ion{H}{1} absorbers in the b/g quasar's spectra of PQ10 and PQ11 and
those of the other 10 quasars.  Possible origins of the large column
densities in the former quasars includes i) there exist over density
regions that is exposed to the same level of ionizing radiation as
those around the other f/g quasars and ii) there exist normal CGM/IGM
absorbers but less illuminated by f/g quasars.  However, the second
scenario is immediately rejected, since the enhancement of the
\ion{H}{1} ionization rate $(1+\omega_{\rm r})$ of PQ10 and PQ11 (14.2
and 7.2) is comparable to those around the other quasar
pairs. Moreover, we do not find any correlations of $\log N_{\rm HI}$
as functions of $R_{\perp}$ and $(1+\omega_{\rm r})$, as seen in the
second and third panels from the left in Figure~\ref{fig:correlation},
though the statistical significance is not so high due to the small
sample size. Only PQ2 has an ionization enhancement ((1+$\omega_{\rm
  r}$) = 205) larger than 100, and much larger than the other quasar
pairs by almost one order of magnitude, but its \ion{H}{1} column
density is comparable to the others.  It should be noted that two f/g
quasars (i.e., PQ10 and PQ11) with optically thick absorbers around
them tend to have larger redshift ($z_{\rm f/g}$ $>$ 2.4).  We will
discuss these in Section~\ref{sec:evolution}.

%%% Figure 7 %%%

\subsection{Stacked Spectrum}
\begin{figure*}[ht!]
  \begin{center}
    \includegraphics[height=8cm,angle=0]{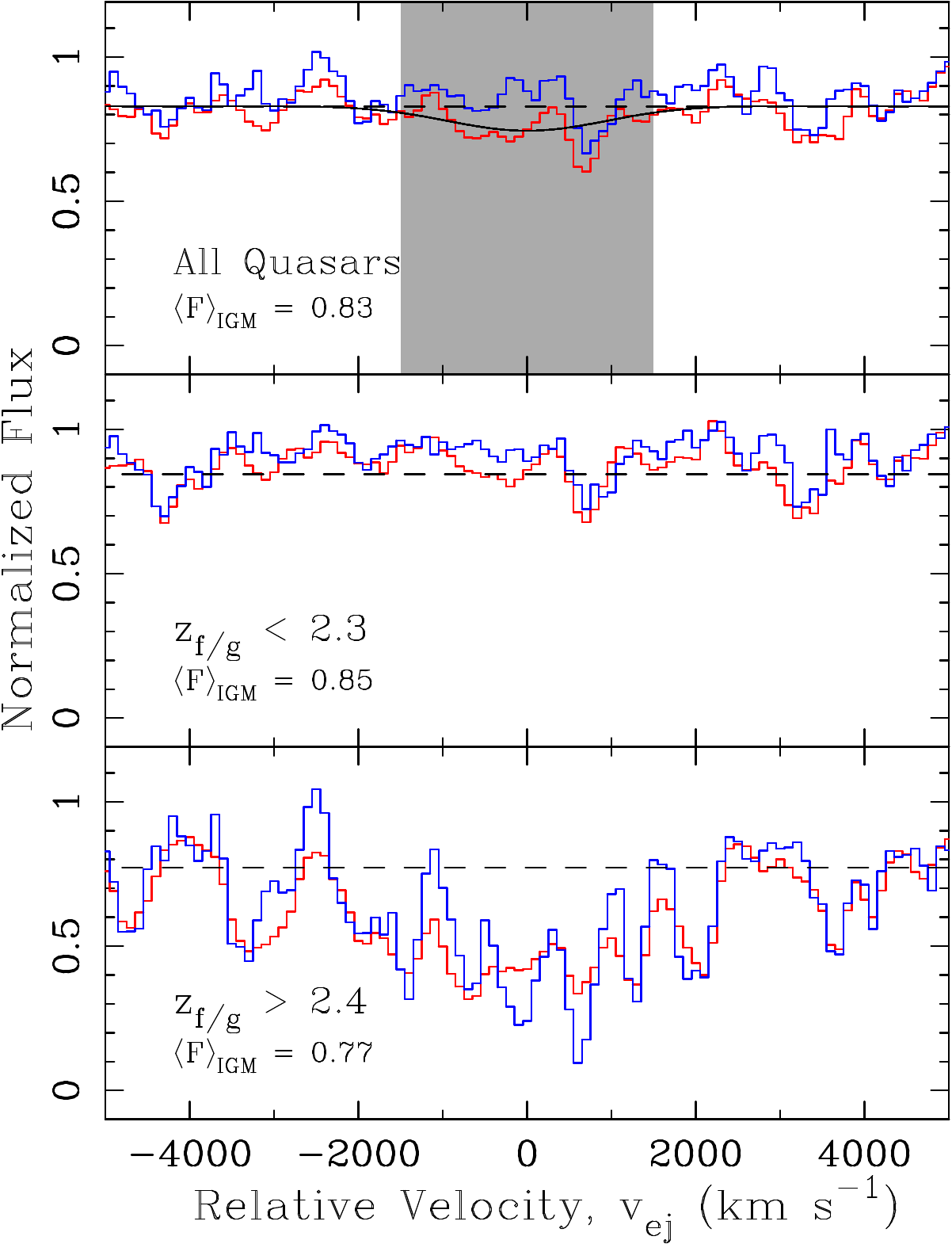}
    \hspace{1cm}
    \includegraphics[height=8cm,angle=0]{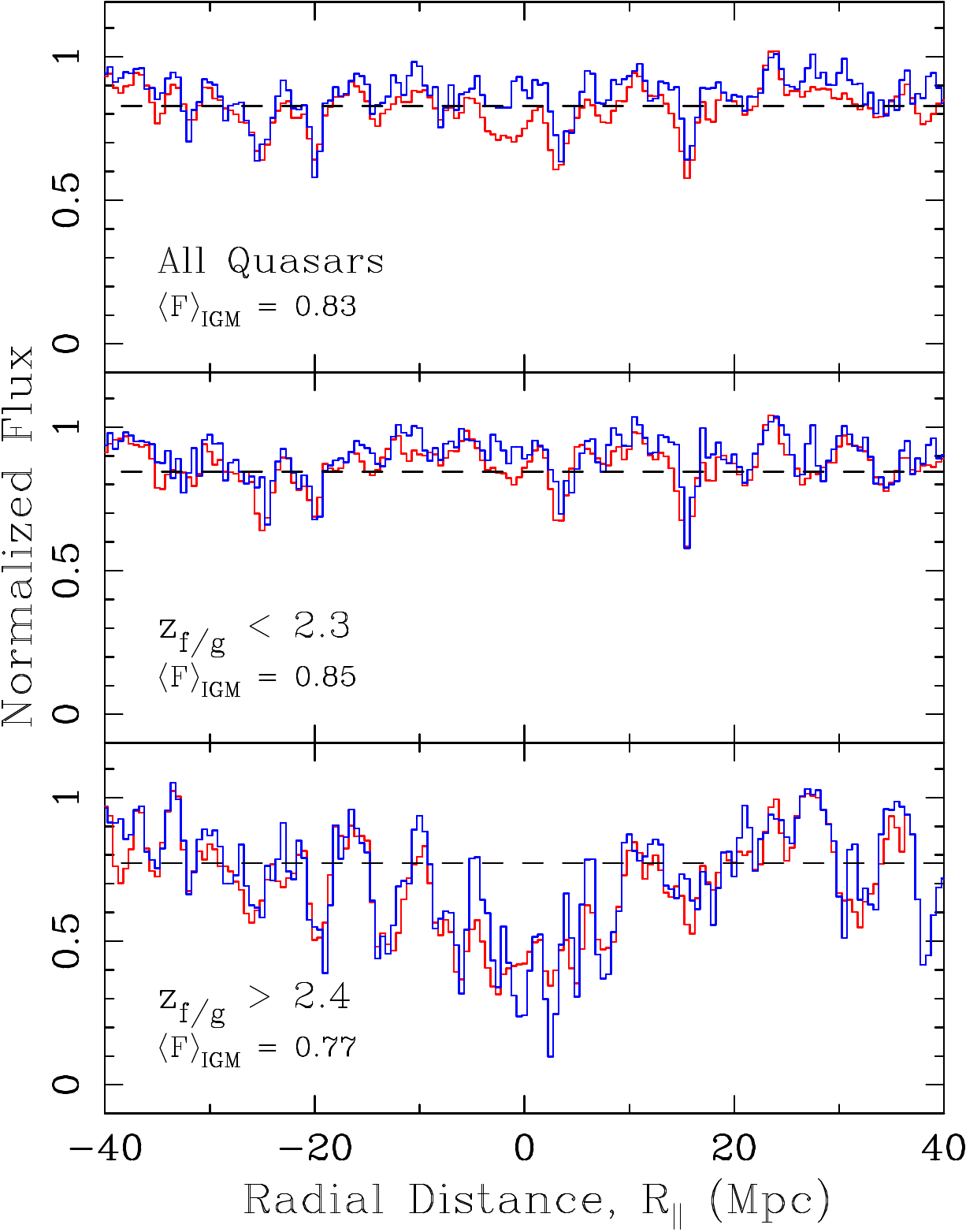}
  \end{center}  
  \caption{Mean (red) and median (blue) spectra of quasars as a
    function of relative velocity (left) or radial distance (right)
    from $z_{\rm f/g}$. Top panel includes all 12 quasars, while
    middle and bottom panels contain only 9 and 3 quasars whose f/g
    quasars at $z_{\rm f/g} < 2.3$ or $z_{\rm f/g} > 2.4$. Horizontal
    dashed lines denote the mean flux of the IGM estimated by
    \citet{fau08}. Solid black curve in a top panel of left figure is
    the mean absorption profile around non-BAL quasars with EW$_{\rm
      rest}$ = 0.99\AA\ and the velocity dispersion of $\sigma$ =
    917~\kms\ at similar perpendicular distance \citep{pro13}. Shaded
    area in the same panel denotes the velocity range (i.e.,
    $\pm$1500~\kms) at which EW$_{\rm rest}$ is measured for our
    stacked spectra.\label{fig:comb_spec1}}
\end{figure*}

\begin{figure}[ht!]
  \begin{center}
    \includegraphics[width=8cm,angle=0]{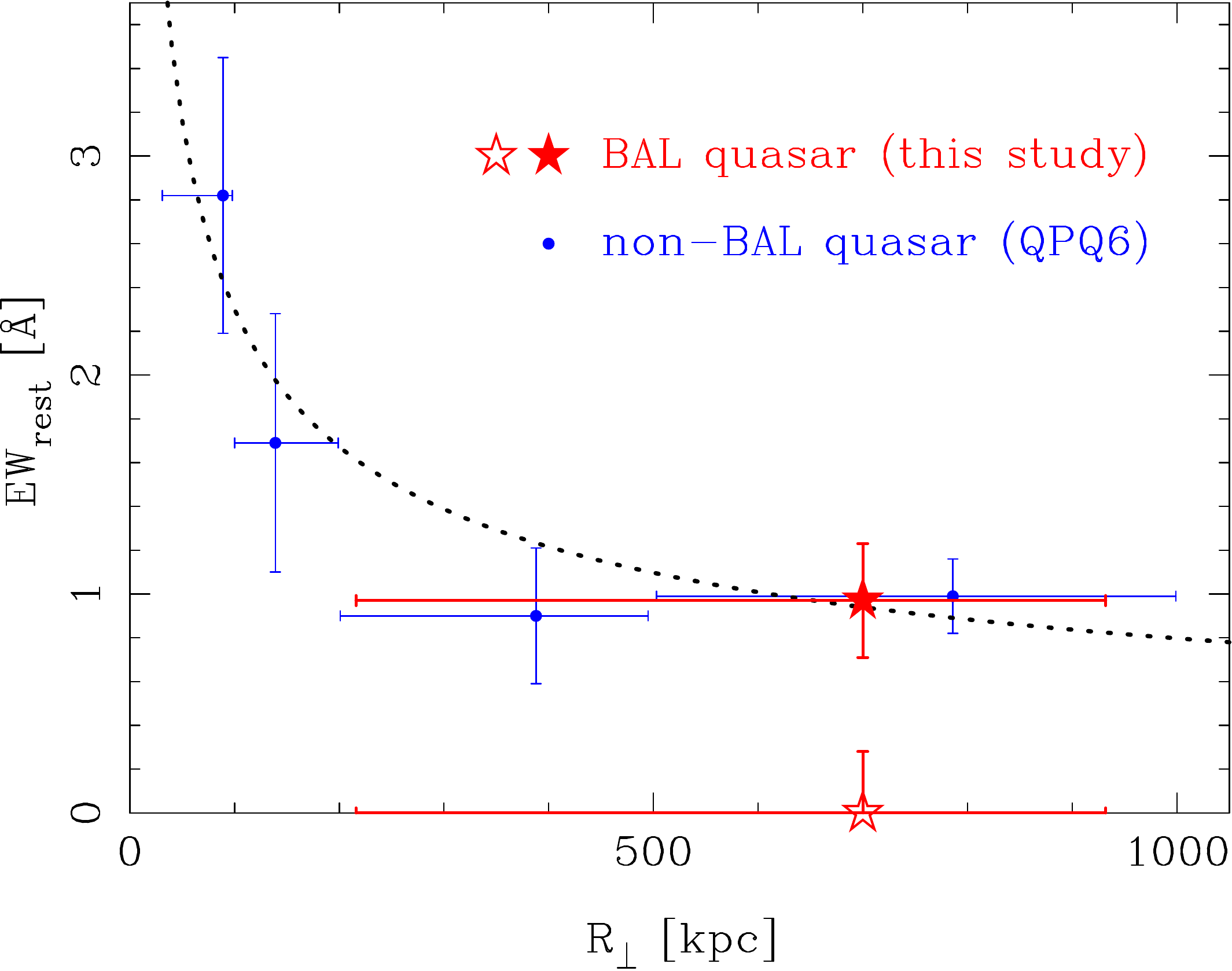}
  \end{center}  
  \caption{Rest-frame equivalent width as a function of the transverse
    distance from f/g quasar. Red filled and open stars with error
    bars are equivalent widths of \ion{H}{1} absorption profile in the
    mean and median stack spectra of our 12 BAL quasars, which denote
    upper and lower limits of EW$_{\rm rest}$. The horizontal error is
    the distribution of $R_{\perp}$ of the 12 quasar pairs, while the
    vertical error is the statistical uncertainty of EW$_{\rm rest}$
    assuming the S/N ratio of the combined spectra ($\sim$8 per
    pixel).  Blue dots with error bars denote EW$_{\rm rest}$ of
    non-BAL quasars \citep{pro13}.  Black dotted curve is a fitted
    model of non-BAL quasars that was introduced by \citet{pro13} as
    ${\rm EW}_{\rm rest} = 2.3 \times \left(R_{\perp}/100~{\rm
      kpc}\right)^{-0.46}$~\AA.\label{fig:ew_measure}}
\end{figure}

It is not easy to detect \ion{H}{1} absorption lines corresponding to
the CGM/IGM around the f/g quasars, since they are severely blended
with \lya\ forest arising in cosmologically distributed IGM.
Therefore, we create the composite spectra of the 12 b/g quasars as a
function of relative velocity or radial distance (assuming Hubble
flow) from $z_{\rm f/g}$, to increase S/N ratio in the stacked
spectrum and to suppress the contamination by unrelated \lya\ forest.
The spectra combined by average and median values are shown in
Figure~\ref{fig:comb_spec1}. After normalizing their flux levels by
those expected from the IGM \citep{fau08}, we measure EW$_{\rm rest}$
of \lya\ absorption directly from the stacked spectra in
$\pm$1500~\kms\ of the center without applying Gaussian fit.  We
marginally detect a weak absorption profile in the mean stack spectrum
with EW$_{\rm rest}$ = 0.97$\pm$0.26\AA\ (which is consistent to the
expected value (EW$_{\rm rest}$ = 0.99$\pm$0.17\AA) from those around
non-BAL quasars \citep{pro13}, as compared in
Figure~\ref{fig:ew_measure}).  However, the mean stack spectrum is
biased significantly due to the existence of the two DLA
systems. Therefore, we also measure EW$_{\rm rest}$ in the median
stack spectrum and acquire EW$_{\rm rest}$ $<$ 0.28\AA.  We will
consider the former and the latter as the upper and lower limits of
EW$_{\rm rest}$.

%%% Figure 8 %%%

%%% Figure 9 %%%

\begin{figure*}[ht!]
  \begin{center}
    \includegraphics[height=8cm,angle=0]{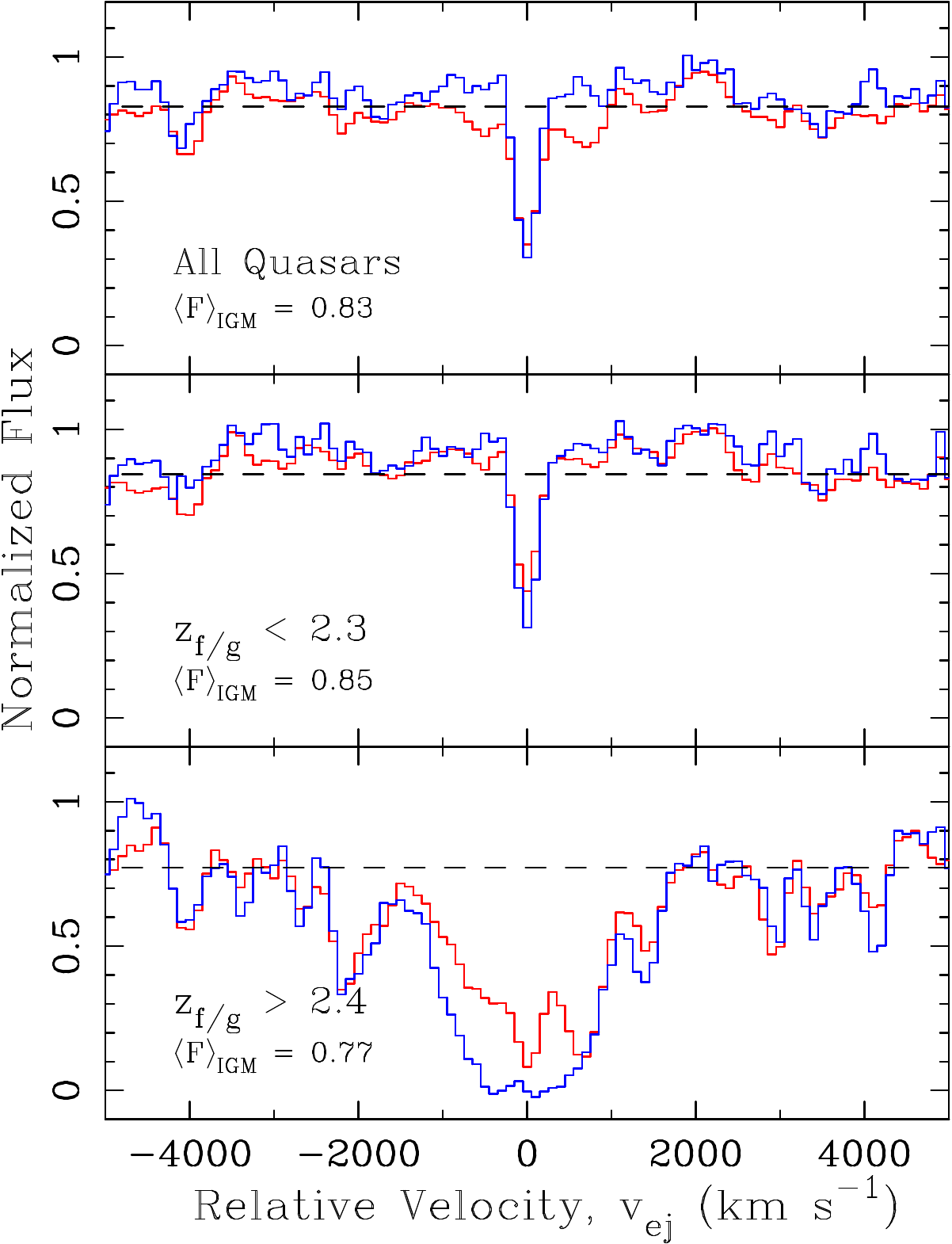}
    \hspace{1cm}
    \includegraphics[height=8cm,angle=0]{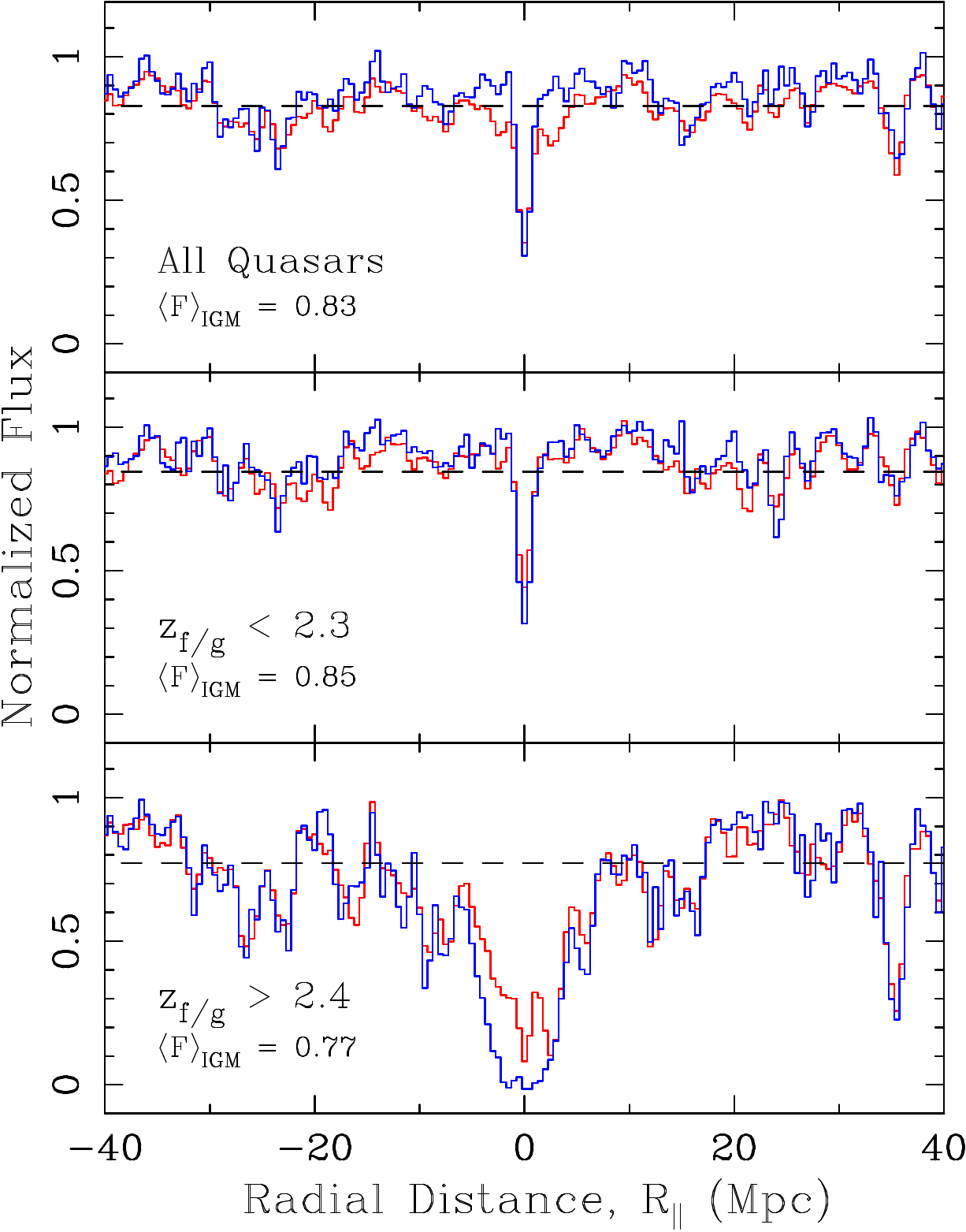}
  \end{center}  
  \caption{Same as Figure~\ref{fig:comb_spec1}, but combined spectra
    after setting their local absorption peaks in $\pm$1500~\kms\ of
    $z_{\rm f/g}$ as the system center.\label{fig:comb_spec2}}
\end{figure*}

We also create composite spectra, using only nine and three b/g
quasars whose corresponding f/g quasars are at $z_{\rm f/g}$ $<$ 2.3
and $>$ 2.4, since f/g quasars with optically thick absorbers (i.e.,
DLA systems) are detected only at $z$ $>$ 2.4.  We do not detect any
remarkable absorption troughs in the combined spectrum of 9 quasars at
lower redshift.  On the other hand, there exists a broad
($\geq$2000~\kms) and strong absorption feature in the combined
spectra of 3 quasars at higher redshift, since two of the three
quasars have optically thick absorption lines as described in
Section~\ref{sec:rate}.  We also create the same stacked spectra after
setting centers of the strongest \ion{H}{1} absorption lines as the
system centers (Figure~\ref{fig:comb_spec2}). In this case, we detect
narrow and deep lines at the centers in both mean and median spectra
combining all 12 quasars and 9 quasars at lower redshift. In the
combined spectrum of 3 higher-$z$ quasars, we detect strong \ion{H}{1}
absorption lines again, but their profiles are deeper and smoother.

%%% Figure 10 %%%

\section{Discussion} \label{sec:discussion}
In this study, we examine the strength of \ion{H}{1} absorption lines
arising in the CGM/IGM around BAL quasars in an impact parameter of
$\lesssim$1~Mpc. We detect optically thick absorbers around two out of
12 BAL quasars whose detection rate is consistent to the past results
around non-BAL quasars \citep{hen06, pro13}.  In the mean stack
spectrum of 12 BAL quasars, we detect a weak \ion{H}{1} absorption
line whose strength is also comparable to that around non-BAL quasars.
However, the statistical confidence in these results is not very high
due to the small sample size. The column densities of optically thick
\ion{H}{1} absorbers around BAL (this study) and non-BAL quasars
\citep{pro13} also have a large difference by about three orders of
magnitude.  In this section, we first discuss if BAL quasars can be
used as an indicator that they are observed from a direction close to
the side. We then consider a possible redshift evolution of BAL
quasars, and finally discuss possible origins of anisotropy of the
proximity effect (between LPE and TPE) around quasars.

\subsection{Inclination Scenario of BAL Quasars} \label{sec:inclination}
The dust torus, an essential structure for the AGN unification model,
has been considered as an origin of the anisotropic \ion{H}{1}
absorption around quasars \citep{pro13,jal19} through shielding effect
of ionizing photon flux radiation from the continuum source. If this
is the case, we expect a small amount of neutral Hydrogen in the
direction with small inclination angles (i.e., close to the face-on
direction from the accretion disk) since they are not shielded by the
dust torus.  It is widely believed that BAL profiles are observed if
we observe the continuum source from near the side since the AGN
outflow wind (i.e., BAL absorber) are accelerated in a direction with
larger inclination angles (e.g., \citealt{mur95,pro00}).  For example,
\citet{elv00} proposes a funnel-shaped thin outflow wind as a BAL
absorber with a cone angle of $\sim$60~degree to the accretion disk
axis.  Thus, {\it positive} (i.e., more ionized) transverse proximity
effect is expected around BAL quasars, while {\it negative} (i.e.,
less ionized) TPE was observed around non-BAL quasars.

Contrary to the expectations above, we found that the detection rate
of optically thick absorbers around BAL quasars in the transverse
direction is at the same level as that seen around non-BAL quasars
($C_{\rm f}$ $\sim$ 0.2). If we consider only DLAs with $\log N_{\rm
  HI}$/(cm$^{-2}$) $\gtrsim$ 20, the detection rate is even larger
around BAL quasars ($C_{\rm f}$ = $0.17^{+0.22}_{-0.11}$) than that
around non-BAL quasars ($C_{\rm f}$ = $0.02 \pm 0.01$;
\citealt{pro13}). The absorption strength (i.e., equivalent width) of
\ion{H}{1} absorption trough in the mean stack spectrum of all 12 BAL
quasars is also consistent to the value that is expected from non-BAL
quasars (see Figure~\ref{fig:ew_measure}).  These results seem to
suggest that BAL quasars may not be a good indicator that they are
observed from a direction close to the dust torus.  Even if the
inclination angles of BAL quasars are indeed large, the dust torus
around them may not be a dominant source of the anisotropic proximity
effect.  Recently, \citet{giu19} propose a global view of AGN
accretion disk winds, in which the outflow winds have smaller
inclination angle relative to the rotation axis with increasing
Eddington ratio, and that they eventually become almost polar-winds in
the case of super-Eddington.  Thus, if this is the case, we need other
parameters of quasars such as Eddington ratio and black hole mass in
addition to the existence of BAL profiles, to examine the effect of
the inclination angle of quasars more precisely.

On the other hand, if we remove DLA systems from our sample, the
detection rate of optically thick absorbers would be $C_{\rm f}$ =
0/10 $\sim$ $0^{+0.18}_{-0}$.  The \ion{H}{1} absorption strength in
the median stack spectrum also becomes smaller (EW$_{\rm rest}$ $<$
0.28\AA).  Although the confidence level is not very high, both of
them are smaller than those around non-BAL quasars, which is
consistent with the anisotropic scenario due to the dust torus. We
will discuss a possible origin of DLA systems around BAL quasars in
Section~\ref{sec:anisotropy}.

\subsection{Redshift Evolution of BAL Quasars} \label{sec:evolution}
As noted in Section~\ref{sec:correlations}, DLAs are detected only
around BAL quasars at higher redshift ($z_{\rm f/g}$ $>$ 2.4) in our
sample.  If we consider only 3 BAL quasars at $z_{\rm f/g}$ $>$ 2.4,
the detection rate of DLA absorber is quite large ($C_{\rm f}$ =
0.67$^{+0.33}_{-0.43}$), while the detection rate is very small
($C_{\rm f}$ = 0$^{+0.21}_{-0}$) at $z_{\rm f/g}$ $<$ 2.4. On the
other hand, around {\it non-BAL} quasars, we do not find any
differences between the detection rates at $z_{\rm f/g}$ $>$ 2.4
($C_{\rm f}$ = 0.02$\pm$0.01) and at $z_{\rm f/g}$ $<$ 2.4 ($C_{\rm
  f}$ = 0.02$\pm$0.01), using the sample of \citet{pro13}.

The difference between BAL quasars at lower and higher redshift could
be related to the evolution effect. For example, the fraction of BAL
quasars increases by a factor of $\sim$3.5 from $z$ $<$ 2.3 to $z$ $>$
3.5, although there is a large uncertainty at $z$ = 2.3 -- 3.0
\citep{all11}.  It is also suggested that quasars with BAL (especially
LoBAL) are probably in the early stage of AGN evolution between ultra
luminous infrared galaxies (ULIRGs) and normal unobscured quasars,
since they have high star formation activity (e.g.,
\citealt{bec00,far07}).

Thus, we speculate that our results could support that BAL
quasars evolve with redshift in terms of ionizing photon flux
radiation; e.g., BAL quasars at higher redshift have not blown off
dust envelope completely and they are still covered by shielding
materials with larger covering fraction like LoBAL quasars.

If this is the case, the two BAL quasars with DLA systems around them
should be highly reddened by dust envelopes, at least compared to the
other 10 BAL quasars without DLAs. To verify this, we perform the
spectral fitting for all f/g BAL quasars assuming the same SED as used
in Section~\ref{sec:enhance} with the extinction law of \citet{cal00}
to estimate the dust reddening.\footnote{Before the fitting, we
  correct for the Galactic extinction using the Galactic color excess
  $E(B-V)$ \citep{sch98} and the extinction law of \citet{car89}.} The
average value of the derived intrinsic color excess of the two BAL
quasars with DLAs is $E(B-V)$ = 0.12$\pm$0.11, while that of the other
10 BAL quasars without DLAs is $E(B-V)$ = 0.33$\pm$0.25. Thus, we do
not find any obvious trends of more dust around the former quasars.
We need to increase the sample size to place more stringent
constraints on redshift evolution of BAL quasars at a statistically
significant level.

\subsection{Anisotropy of Proximity Effect around Quasars} \label{sec:anisotropy}
The anisotropic proximity effect cannot be reconciled with an
isotropically emitting flux source.  In addition to the enhancement of
ionizing flux radiation, \citet{jal19,jal21} also considered the gas
over-density effect around quasars. They derived the excess over
density up to $\sim$5~Mpc from quasars in the both line-of-sight and
transverse directions, but they confirmed that the material in the
transverse direction should not be illuminated more than 27\%\ of the
radiation level in the line-of-sight direction from the quasars to
reconcile the difference (i.e., the anisotropic proximity effect still
remains even after considering gas over density.).

There are several interpretations for the anisotropy such as i) an
intrinsic anisotropy of the ionizing flux radiation due to structures
of host galaxies and/or the accretion disks, ii) anisotropic
obscuration by the dust torus, iii) the finite lifetime of the
quasars\footnote{Quasar lifetime is short enough for ionizing photons
  from the f/g quasars to propagate into the CGM/IGM along the lines
  of sight to b/g quasars ($t_{\rm qso}$ $\lesssim$ 1~Myr). Since a
  typical scale of LPE and TPE studies is $\sim$1~Mpc (i.e., millions
  of light years), the light travel time from the CGM/IGM in the
  line-of-sight direction (after being sufficiently ionized) is
  shorter than that from the CGM/IGM in the transverse direction
  (before being ionized) by a few Myrs.}, and iv) the inflow of matter
into (or the outflow of the matter from) the halo of quasar host
galaxies (e.g., \citealt{pro13,jal19}, and references therein).

In this study, we verified the anisotropic obscuration scenario by the
dust torus as the origin of the difference between LPE and TPE. This
scenario has been thought to be the most probable one, since the
existence of the dust torus has been confirmed by various observations
(e.g., \citealt{ant93}) and its opening angle is constrained to be
large enough \citep{sch04,kir08,lu11}. Contrary to the expectation, we
detect i) two optically thick DLA-like absorbers and ii) a weak
\ion{H}{1} absorption in the mean stack spectrum with EW$_{\rm rest}$
$\sim$ 1\AA\ (at the same level as non-BAL quasars) around BAL quasars
whose inclination angle is expected to be large (i.e., close to the
edge-on along our lines of sight).

However, the column densities of optically thick absorbers differ by
about three orders of magnitude between those around BAL quasars
(i.e., DLA-like absorbers with $\log N_{\rm HI}$/(cm$^{-2}$) $\sim$
20; this study) and non-BAL quasars (i.e., Lyman limit system
(LLS)-like absorbers with $\log N_{\rm HI}$/(cm$^{-2}$) $\gtrsim$
17.2; \citealt{pro13}).  Indeed, around non-BAL quasars, both the
covering fraction ($C_{\rm f}$ $\sim$ 0\%\ at $R_{\perp}$ $\sim$
1~Mpc) and the clustering amplitude ($r_{\rm 0}$ $\sim$ 4~Mpc) of the
quasar-absorber cross-correlation function\footnote{$\xi_{\rm QA}(r) =
  (r/r_{\rm 0})^{-\gamma}$, where $\gamma$ = 1.6, assuming
  galaxy-galaxy clustering \citep{ade05}.}  of DLAs is very different
from those of LLSs ($C_{\rm f}$ $\sim$ 20\%\ and $r_{\rm 0}$ $\sim$
19~Mpc), which suggests that the former traces galaxies within the
same dark matter halos as f/g BAL quasars and that the latter arises
at a dense, self-shielding media in large-scale structures like
filaments \citep{pro13}.  If this is the case, the statistical
analyses based on the large sample size of the systems without
DLA-like systems would be required to probe the environment around the
BAL quasars since the existence of galaxies would bias toward larger
column densities.  As noted in Section~\ref{sec:inclination}, the
covering fraction of optically thick gas around BAL quasars (after
removing DLA-like systems) would be smaller than that around non-BAL
quasars although not statistically significant, which is expected from
the anisotropic scenario \citep{pro13,jal19}.

\section{Summary} \label{subsec:summary}
In this study, we examined environments of BAL quasars whose
inclination angle is considered to be large (i.e., we observe them
close to the side.).  To study \ion{H}{1} absorption strength around
foreground BAL quasars in spectra of background quasars, we targeted
closely separated 12 quasar pairs at different redshift with
separation angle of $\theta$ $<$ 120$^{\prime\prime}$
($\lesssim$~1~Mpc at $z$ $\sim$ 2).  We used low-resolution spectra
($R$ $\sim$ 2000) of SDSS quasars and those obtained by ourselves with
Subaru/FOCAS.  Because it is difficult to fit the continuum level
directly in the spectral region of \lya\ forest, we first roughly
estimated the quasar's intrinsic spectra using a PCA method with
template spectra of low-redshift quasars. Then, we manually normalized
them again with a low-order spline function to satisfy the mean flux
of the IGM.

We measured the \ion{H}{1} absorption strength in an individual quasar
spectrum.  We confirmed there exist optically thick gas around two BAL
quasars, whose detection rate is consistent to the past results around
non-BAL quasars.  We also created the composite mean spectrum of all
12 background quasars and marginally detected a weak absorption
profile at $z_{\rm f/g}$, whose absorption strength is consistent with
the expected value from those around non-BAL quasars.

We also confirmed that only high redshift BAL quasars at $z_{\rm f/g}$
$>$ 2.4 have the corresponding optically thick gas, which implies a
possible redshift evolution of BAL quasars. However, we do not find
any obvious trends of more dust around higher redshift BAL quasars
based on our relatively small sample size.

Taking the results above into account, the anisotropic obscuration due
to the dust torus cannot reconcile the difference in the \ion{H}{1}
absorption strength around quasars by itself. However, these results
are not statistically significant due to the small sample size.
Moreover, if we remove DLAs (whose origin is probably different from
those of LLSs) from our sample, both the covering fraction and the
mean absorption strength of optically thick gas around BAL quasars
would be smaller than that around non-BAL quasars, which still
supports the anisotropic scenario.

There are other possible ideas for the anisotropic proximity effect.
For example, the quasars may have a finite lifetime and their ionizing
photons have not reached to the CGM/IGM in the transverse direction.
Another possibility is that the two BAL quasars with DLA systems
(i.e., PQ10 and PQ11) have smaller inclination angle than those of the
other BAL quasars so that the CGM/IGM in the transverse direction are
less ionized like those around non-BAL quasars.

For narrowing down possible scenarios further, we plan to increase the
sample size significantly by searching for fainter quasar pairs that
satisfy all the criteria except for $g$-band magnitude in
Section~\ref{sec:sample}, using Prime Focus Spectrograph that will be
installed to Subaru telescope soon.

\begin{acknowledgments}
We thank the anonymous referee for very useful comments and
suggestions. This research is based in part on data collected at the
Subaru Telescope, which is operated by the National Astronomical
Observatory of Japan. We are honored and grateful for the opportunity
of observing the Universe from Maunakea, which has the cultural,
historical, and natural significance in Hawaii. This work was
supported by JSPS KAKENHI Grant Number 21H01126.

Funding for the Sloan Digital Sky Survey IV has been provided by the
Alfred P. Sloan Foundation, the U.S.  Department of Energy Office of
Science, and the Participating Institutions. SDSS-IV acknowledges
support and resources from the Center for High Performance Computing
at the University of Utah. The SDSS website is www.sdss.org.

SDSS-IV is managed by the Astrophysical Research Consortium for the
Participating Institutions of the SDSS Collaboration including the
Brazilian Participation Group, the Carnegie Institution for Science,
Carnegie Mellon University, Center for Astrophysics | Harvard \&
Smithsonian, the Chilean Participation Group, the French Participation
Group, Instituto de Astrof\'isica de Canarias, The Johns Hopkins
University, Kavli Institute for the Physics and Mathematics of the
Universe (IPMU) / University of Tokyo, the Korean Participation Group,
Lawrence Berkeley National Laboratory, Leibniz Institut f\"ur
Astrophysik Potsdam (AIP), Max-Planck-Institut f\"ur Astronomie (MPIA
Heidelberg), Max-Planck-Institut f\"ur Astrophysik (MPA Garching),
Max-Planck-Institut f\"ur Extraterrestrische Physik (MPE), National
Astronomical Observatories of China, New Mexico State University, New
York University, University of Notre Dame, Observat\'ario Nacional /
MCTI, The Ohio State University, Pennsylvania State University,
Shanghai Astronomical Observatory, United Kingdom Participation Group,
Universidad Nacional Aut\'onoma de M\'exico, University of Arizona,
University of Colorado Boulder, University of Oxford, University of
Portsmouth, University of Utah, University of Virginia, University of
Washington, University of Wisconsin, Vanderbilt University, and Yale
University.
\end{acknowledgments}


\clearpage
\begin{thebibliography}{}
\bibitem[Adelberger et al.(2005)]{ade05} Adelberger, K.~L., Steidel,
  C.~C., Pettini, M., et al.\ 2005, \apj, 619, 697. doi:10.1086/426580
\bibitem[Allen et al.(2011)]{all11} Allen, J.~T., Hewett, P.~C.,
  Maddox, N., et al.\ 2011, \mnras, 410,
  860. doi:10.1111/j.1365-2966.2010.17489.x
\bibitem[Antonucci(1993)]{ant93} Antonucci, R.\ 1993, \araa, 31,
  473. doi:10.1146/annurev.aa.31.090193.002353
\bibitem[Bajtlik et al.(1988)]{baj88} Bajtlik, S., Duncan, R.~C., \&
  Ostriker, J.~P.\ 1988, \apj, 327, 570. doi:10.1086/166217
\bibitem[Becker et al.(2000)]{bec00} Becker, R.~H., White, R.~L.,
  Gregg, M.~D., et al.\ 2000, \apj, 538, 72. doi:10.1086/309099
\bibitem[Cardelli et al.(1989)]{car89} Cardelli, J.~A., Clayton,
  G.~C., \& Mathis, J.~S.\ 1989, \apj, 345, 245. doi:10.1086/167900
\bibitem[Calzetti et al.(2000)]{cal00} Calzetti, D., Armus, L.,
  Bohlin, R.~C., et al.\ 2000, \apj, 533, 682. doi:10.1086/308692
\bibitem[Chen et al.(2022)]{che22} Chen, Z., He, Z., Ho, L.~C., et
  al.\ 2022, Nature Astronomy, 6, 339. doi:10.1038/s41550-021-01561-3
\bibitem[Croft(2004)]{cro04} Croft, R.~A.~C.\ 2004, \apj, 610,
  642. doi:10.1086/421839
\bibitem[Elvis(2000)]{elv00} Elvis, M.\ 2000, \apj, 545,
  63. doi:10.1086/317778
\bibitem[Farrah et al.(2007)]{far07} Farrah, D., Lacy, M., Priddey,
  R., et al.\ 2007, \apjl, 662, L59. doi:10.1086/519492
\bibitem[Faucher-Gigu{\`e}re et al.(2008)]{fau08} Faucher-Gigu{\`e}re,
  C.-A., Prochaska, J.~X., Lidz, A., et al.\ 2008, \apj, 681,
  831. doi:10.1086/588648
\bibitem[Filiz Ak et al.(2013)]{fil13} Filiz Ak, N., Brandt, W.~N.,
  Hall, P.~B., et al.\ 2013, \apj, 777,
  168. doi:10.1088/0004-637X/777/2/168
\bibitem[Gehrels(1986)]{geh86} Gehrels, N.\ 1986, \apj, 303,
  336. doi:10.1086/164079
\bibitem[Giustini \& Proga(2019)]{giu19} Giustini, M. \& Proga,
  D.\ 2019, \aap, 630, A94. doi:10.1051/0004-6361/201833810
\bibitem[Haardt \& Madau(2012)]{har12} Haardt, F. \& Madau, P.\ 2012,
  \apj, 746, 125. doi:10.1088/0004-637X/746/2/125
\bibitem[Hall et al.(2002)]{hal02} Hall, P.~B., Anderson, S.~F.,
  Strauss, M.~A., et al.\ 2002, \apjs, 141, 267. doi:10.1086/340546
\bibitem[Hennawi et al.(2006)]{hen06} Hennawi, J.~F., Prochaska,
  J.~X., Burles, S., et al.\ 2006, \apj, 651, 61. doi:10.1086/507069
\bibitem[Hennawi \& Prochaska(2013)]{hen13} Hennawi, J.~F. \&
  Prochaska, J.~X.\ 2013, \apj, 766,
  58. doi:10.1088/0004-637X/766/1/58
\bibitem[Ishimoto et al.(2020)]{ish20} Ishimoto, R., Kashikawa, N.,
  Onoue, M., et al.\ 2020, \apj, 903, 60. doi:10.3847/1538-4357/abb80b
\bibitem[Ishita et al.(2021)]{ish21} Ishita, D., Misawa, T., Itoh, D.,
  et al.\ 2021, \apj, 921, 119. doi:10.3847/1538-4357/ac14b4
\bibitem[Jalan et al.(2019)]{jal19} Jalan, P., Chand, H., \& Srianand,
  R.\ 2019, \apj, 884, 151. doi:10.3847/1538-4357/ab4191
\bibitem[Jalan et al.(2021)]{jal21} Jalan, P., Chand, H., \& Srianand,
  R.\ 2021, \mnras, 505, 689. doi:10.1093/mnras/stab1303
\bibitem[Khaire \& Srianand(2015)]{kha15} Khaire, V. \& Srianand,
  R.\ 2015, \apj, 805, 33. doi:10.1088/0004-637X/805/1/33
\bibitem[Kirkman \& Tytler(2008)]{kir08} Kirkman, D. \& Tytler,
  D.\ 2008, \mnras, 391, 1457. doi:10.1111/j.1365-2966.2008.13994.x
\bibitem[Knigge et al.(2008)]{kni08} Knigge, C., Scaringi, S., Goad,
  M.~R., et al.\ 2008, \mnras, 386,
  1426. doi:10.1111/j.1365-2966.2008.13081.x
\bibitem[L{\'\i}pari \& Terlevich(2006)]{lip06} L{\'\i}pari, S.~L. \&
  Terlevich, R.~J.\ 2006, \mnras, 368,
  1001. doi:10.1111/j.1365-2966.2006.10215.x
\bibitem[Lu \& Yu(2011)]{lu11} Lu, Y. \& Yu, Q.\ 2011, \apj, 736,
  49. doi:10.1088/0004-637X/736/1/49
\bibitem[Lusso et al.(2015)]{lus15} Lusso, E., Worseck, G., Hennawi,
  J.~F., et al.\ 2015, \mnras, 449, 4204. doi:10.1093/mnras/stv516
\bibitem[Lyke et al.(2020)]{lyk20} Lyke, B.~W., Higley, A.~N., McLane,
  J.~N., et al.\ 2020, \apjs, 250, 8. doi:10.3847/1538-4365/aba623
\bibitem[Murray et al.(1995)]{mur95} Murray, N., Chiang, J., Grossman,
  S.~A., et al.\ 1995, \apj, 451, 498. doi:10.1086/176238
\bibitem[P{\^a}ris et al.(2017)]{par17} P{\^a}ris, I., Petitjean, P.,
  Ross, N.~P., et al.\ 2017, \aap, 597,
  A79. doi:10.1051/0004-6361/201527999
\bibitem[P{\^a}ris et al.(2011)]{par11} P{\^a}ris, I., Petitjean, P.,
  Rollinde, E., et al.\ 2011, \aap, 530,
  A50. doi:10.1051/0004-6361/201016233
\bibitem[Prochaska et al.(2013)]{pro13} Prochaska, J.~X., Hennawi,
  J.~F., Lee, K.-G., et al.\ 2013, \apj, 776,
  136. doi:10.1088/0004-637X/776/2/136
\bibitem[Proga et al.(2000)]{pro00} Proga, D., Stone, J.~M., \&
  Kallman, T.~R.\ 2000, \apj, 543, 686. doi:10.1086/317154
\bibitem[Schirber et al.(2004)]{sch04} Schirber, M.,
  Miralda-Escud{\'e}, J., \& McDonald, P.\ 2004, \apj, 610,
  105. doi:10.1086/421451
\bibitem[Schlegel et al.(1998)]{sch98} Schlegel, D.~J., Finkbeiner,
  D.~P., \& Davis, M.\ 1998, \apj, 500, 525. doi:10.1086/305772
\bibitem[Scott et al.(2000)]{sco00} Scott, J., Bechtold, J.,
  Dobrzycki, A., et al.\ 2000, \apjs, 130, 67. doi:10.1086/317340
\bibitem[Shen et al.(2016)]{she16} Shen, Y., Brandt, W.~N., Richards,
  G.~T., et al.\ 2016, \apj, 831, 7. doi:10.3847/0004-637X/831/1/7
\bibitem[Suzuki et al.(2005)]{suz05} Suzuki, N., Tytler, D., Kirkman,
  D., et al.\ 2005, \apj, 618, 592. doi:10.1086/426062
\bibitem[Telfer et al.(2002)]{tel02} Telfer, R.~C., Zheng, W., Kriss,
  G.~A., et al.\ 2002, \apj, 565, 773. doi:10.1086/324689
\bibitem[Treister \& Urry(2005)]{tre05} Treister, E. \& Urry,
  C.~M.\ 2005, \apj, 630, 115. doi:10.1086/431892
\bibitem[Ueda et al.(2003)]{ued03} Ueda, Y., Akiyama, M., Ohta, K., et
  al.\ 2003, \apj, 598, 886. doi:10.1086/378940
\bibitem[Urrutia et al.(2009)]{urr09} Urrutia, T., Becker, R.~H.,
  White, R.~L., et al.\ 2009, \apj, 698,
  1095. doi:10.1088/0004-637X/698/2/1095
\bibitem[Vanden Berk et al.(2001)]{van01} Vanden Berk, D.~E.,
  Richards, G.~T., Bauer, A., et al.\ 2001, \aj, 122,
  549. doi:10.1086/321167
\bibitem[Weymann et al.(1991)]{wey91} Weymann, R.~J., Morris, S.~L.,
  Foltz, C.~B., et al.\ 1991, \apj, 373, 23. doi:10.1086/170020
\end{thebibliography}
\end{document}